\documentclass[prb,%
 reprint,
 amsmath,amssymb,
 aps,
]{revtex4-2}

\usepackage{graphicx}
\usepackage{dcolumn}
\usepackage{bm} 
\usepackage{xcolor}

\usepackage[hidelinks]{hyperref}

\newcounter{subfigure}

\graphicspath{{Figures/}} 

\begin{document}


\title{Time and Momentum Resolved Tunneling Spectroscopy of Floquet dynamics}

\author{Lucas Q. Silveira}
\affiliation{Physics Department, Northeastern University, Boston, MA 02115, USA}
 
\author{Adrian E. Feiguin}
\affiliation{Physics Department, Northeastern University, Boston, MA 02115, USA}

\date{\today}

\begin{abstract}

    Periodically driven quantum systems provide a powerful route to engineer novel states of matter by controlling their effective Hamiltonians through external fields. However, most treatments are usually simplified by considering the high-frequency limit, and do not account for the processes taking place in the transient regime during the onset of the drive. In this work, we introduce a time and momentum resolved tunneling spectroscopy protocol to probe the instantaneous energy spectrum of Floquet-driven systems beyond the high-frequency regime, capturing both emergent effects and non-adiabatic phenomena. We benchmark the method on driven non-interacting fermionic models, and then generalize the approach to strongly correlated systems with the aid of time-dependent density matrix renormalization group techniques. We also provide a microscopic description to the emergence of in-gap states under resonant driving, and briefly explore its finite temperature analog.
\end{abstract}

\maketitle


\section{Introduction}
\label{sec:Introduction}

Tailoring phases of matter with novel quantum properties has historically been one of the most challenging tasks in condensed matter physics, either because of the fragility of their governing quantum effects or their lack of natural occurrence. However, the study of out-of-equilibrium effects provides an alternative to such obstacles. By fine-tuning the coupling of an initially trivial phase to an external driving source, one can reveal out-of-equilibrium phases that emerge from the interplay between degrees of freedom across different energy scales. Thus, gaining control over the band structure and inducing features such as superconductivity \cite{light-induced2016superconductivity, claassen2019superconductor} and topological orders \cite{rudner2020induced-topology, optical-lattice2014topological, topological2022controlling}. These \textit{non-thermal pathways} \cite{Alberto2021Colloquium} essentially delay relaxation of excited states and steer the system to a pre-thermal phase.

A realization of this approach, referred as \textit{Floquet engineering} \cite{2020FloquetHandbook, Anatoli2015High-frequency, oka2019FloquetEngineering, weitenberg2021tailoring}, has gained significant interest due to the advancements of ultra-fast light sources, which enable a direct observation of quantum dynamics in the femtosecond timescale 
\cite{Alberto2021Colloquium}. The idea is that by periodically driving the system away from equilibrium with a time-dependent external source (e.g. electromagnetic fields) electronic states become dressed and suppression or enhancement of specific interactions in the Hamiltonian can be selected. In addition to that, a perturbative expansion over the driving frequency can be used to engineer interaction terms, which were not previously present in the original Hamiltonian description. Interestingly, experimental observation of these effects can be obtained by exploiting time-resolved and angle-resolved photoemission spectroscopy (tr-ARPES), which provides both temporal and momentum resolution required to probe these transient regimes \cite{exp1-2023, exp2-2013, exp3-2016}. Nevertheless, as we drive the system away from equilibrium, the imaginary part of the retarded Green's function is not guaranteed to remain positive and, therefore, does not lead to the usual density of states interpretation \cite{uhrig2019positivity}.

To address this theoretical limitation, we improve on the method developed at \cite{Krissia2019, Krissia2020RIXS, Krissia2023core-hole}, where it was shown that the time-resolved energy spectrum can be obtained by means of  \textit{extended tunneling spectroscopy}. The idea is inspired by prior tunneling experiments in one-dimensional geometries \cite{auslaender2002,auslaender2005} and recently generalized to two dimensions in the so-called ``quantum twisting microscope'' \cite{Inbar2023,Birkbeck2025,Xiao2026}. The basic premise is that, by introducing a non-interacting probe parallel to the sample, and applying a bias voltage $V_g$ through it, the allowed tunneling processes between the sample and probe are constrained by momentum and energy conservation. As a result, the energy spectrum can be determined by computing the momentum distribution along the probe for every time step. From a numerical perspective, this approach offers an advantage over conventional techniques, since each computation naturally encodes the information about the available states at given momentum and energy, without prior knowledge of the whole spectrum. In this work, we consider the coupling between sample and probe to follow a Gaussian shape, as opposed to the previously employed step function. With this modification, we are able to properly explore time resolved spectra, since the finite width of the probe allows spectral features between different instants to be discerned. Furthermore, given the main interest of this work is Floquet systems, the sudden turn-on of the probe can lead to undesired spectral leakage, which may interfere with the real spectral features arising from the drive.

This paper is organized as follows. In Sec. \ref{sec:Floquet driven systems}, we review the basic concepts of periodically driven systems and the Floquet theorem. In Sec. \ref{sec:Method}, we introduce the proposed time and momentum resolved tunneling spectroscopy protocol and describe its numerical implementation. In Sec. \ref{sec:Results}, we present results for both non-interacting and interacting systems as they are driven away from equilibrium. We close this section by providing a microscopic description of the emergent in-gap states in a resonantly driven Charge Density Wave (CDW) insulator. A complementary result at finite temperature is also provided. These in-gap states were previously reported numerically in Ref.~\onlinecite{Germans2023InGap}, but their underlying mechanism has remained elusive. In Appendix \ref{sec:AppendixA}, details on the driven non-interacting system and its response under the proposed spectroscopy protocol are explicitly derived. Finally, comments on the derivation of (\ref{t-V chain rotating}) and (\ref{spectral function}) are presented in Appendix \ref{sec:appendixB} and Appendix \ref{AppendixC}, respectively.

\section{Floquet driven systems}
\label{sec:Floquet driven systems}

\begin{figure}
    \includegraphics[width=.8\linewidth]{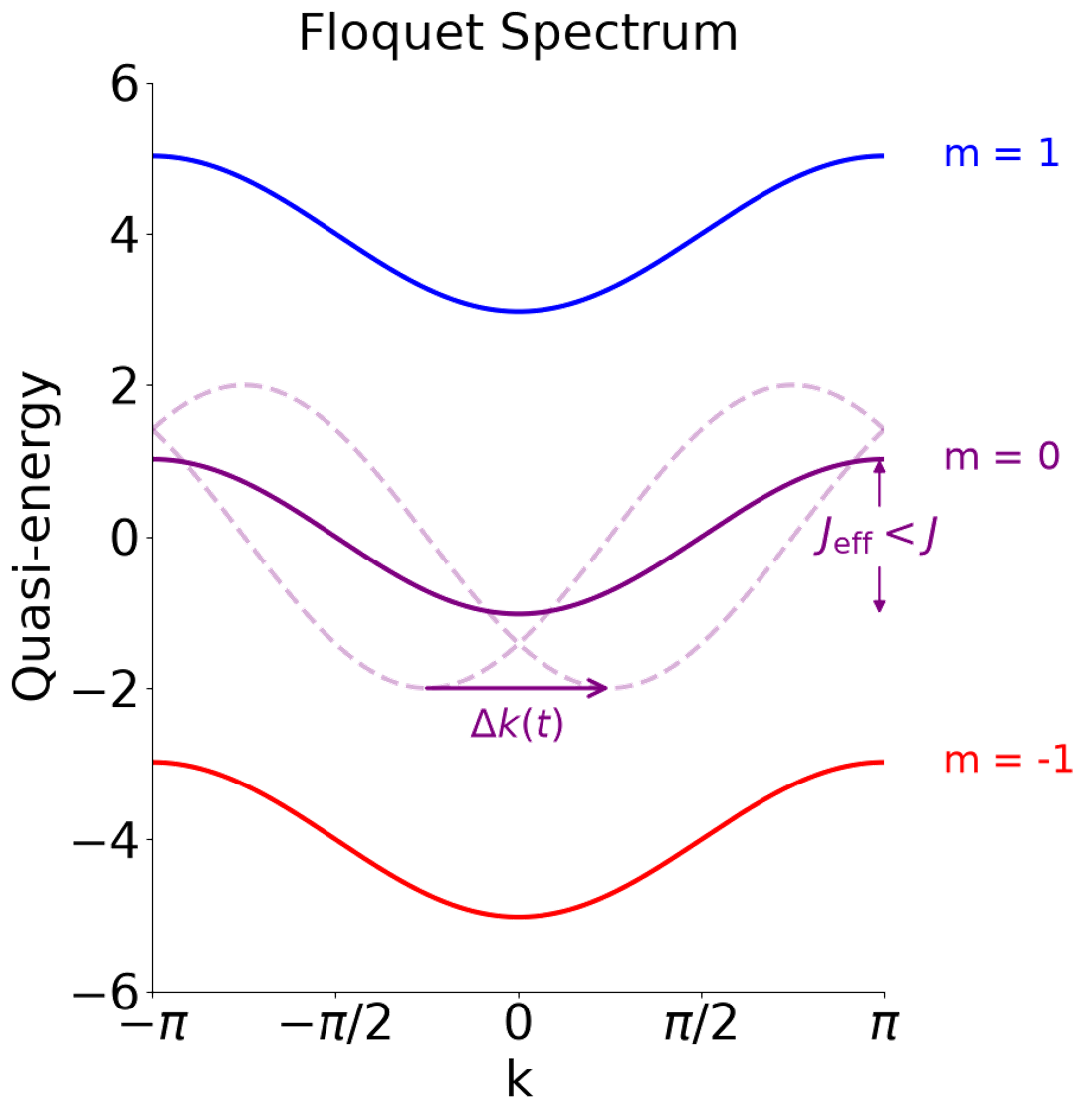}
    \caption{Quasi-energy spectrum of a periodically driven tight-binding chain. Solid lines indicate the different Floquet replicas separated by $\Omega=2\pi/T$. Dashed lines represent the instantaneous energy spectrum as it oscillates over time due to the momentum shift $\Delta k(t)=A_0\sin(\Omega t)$, which in the effective description leads to the renormalization of the bandwidth.}
    \label{fig:theory}
\end{figure}

A result known as \textit{Floquet's theorem} establishes a constraint on the wave function of systems whose Hamiltonians satisfy $H(t+T)=H(t)$. This result follows from the observation that the unitary time-evolution operator admits the decomposition 
\begin{align}\nonumber
    U(t+T, t_0)&=U(t+T, t_0+ T)U(t_0+T, t_0)
    \\
    &=U(t, t_0)U(t_0+T, t_0)
\end{align}
and, therefore, can be understood as a fast non-periodic motion given by $U(t,t_0)$, and a slow time-independent part that repeats itself over a period
\begin{align}\label{U_F}
    U(t_0+T, t_0)=e^{-iH_F[t_0]T}.
\end{align}
Consequently, the wave function is required to be of the form  
\begin{align}
    |\Psi_\alpha(t)\rangle\!=\!e^{-i\varepsilon_\alpha (t-t_0)}|\Phi_\alpha(t)\rangle, 
    \hspace{1mm} 
    |\Phi_\alpha(t+T)\rangle=|\Phi_\alpha(t)\rangle.
\end{align}
It is interesting to note that the spectrum of the so called \textit{Floquet Hamiltonian}, introduced in equation (\ref{U_F}), is defined up to integer multiples of the driving frequency $\Omega=2\pi/T$. This induces the appearance of replicas of the energy bands (essentially electronic states dressed by photons), which can then trigger transitions and hybridization processes otherwise forbidden in the static regime. See Fig. \ref{fig:theory}.

Furthermore, since both the Hamiltonian and the wave-function share the same periodicity, a Fourier expansion allows to recast the Schr\"odinger equation in terms of a time-independent eigenvalue problem coupling the different Fourier modes
\begin{align}
    (\varepsilon_\alpha-m \Omega)|\Phi^{(m)}_\alpha\rangle=\sum_{m^\prime}H^{(m-m^\prime)}|\Phi^{(m^\prime)}_\alpha\rangle,
\end{align}
where we used $H(t)=\sum_m  e^{im\Omega t} H^{(m)}$ as well as $|\Phi_\alpha (t)\rangle=\sum_m e^{-im\Omega t}|\Phi_\alpha^{(m)}\rangle$. 

Note, however, that as $\Omega$ increases, the replicas of the energy band move farther apart in energy. Consequently, the coupling between distinct Fourier modes becomes suppressed, and a high-frequency expansion can be carried out perturbatively. Physically, this is a manifestation of the slow degrees of freedom responding only to the time-averaged effect of the fast drive. Mathematically, this is expressed in terms of the Floquet-Magnus expansion \cite{Anatoli2015High-frequency, oka2019FloquetEngineering}
\begin{align}\label{effective Hamiltonian}
    H_{eff}&=H^{(0)}+\sum_{m=1}^\infty \frac{[H^{(m)},H^{(-m)}]}{m\Omega}+\mathcal{O}(1/\Omega^2).
\end{align}

As previously mentioned, the first term represents the time-averaged dynamics of the system over a period of the drive. This amounts to the renormalization of the existing parameters in the Hamiltonian, thus providing the control over the dominant interactions by tuning the drive's properties. Meanwhile, the $1/\Omega$ correction describes a virtual process in which the system absorbs $m$ photons with energy $\Omega$ and subsequently re-emits them. These photon-assisted transitions effectively couple the high-energy degrees of freedom to the low energy dynamics, which lead to the appearance of interactions not present in the original Hamiltonian.

As an example, we now discuss the phenomenon of \textit{dynamical localization} \cite{Dunlap1986dynamical}. Consider a one-dimensional tight-binding model subjected to a time-periodic vector potential $A(t)=A_0\sin(\Omega t)$
\begin{align}
    H=-J\sum_i \bigg(e^{-iA(t)}c^\dagger_ic_{i+1}+h.c.\bigg),
\end{align}
where we introduce the hopping parameter as $J$, and reserve $t$ for time.
This setup is of special interest once the time-dependent phase admits an exact Fourier decomposition through the Jacobi-Anger identity
\begin{align}\label{Jacobi-Anger}
    e^{-iA_0\sin(\Omega t)}=\sum_{m=-\infty}^{\infty}\mathcal{J}_m(A_0)e^{-im\Omega t},
\end{align}
where $\mathcal{J}_m(A_0)$ denotes the Bessel function of the first kind. Consequently, upon retaining the $m=0$ contribution, the first correction to the energy band becomes
\begin{align}
    \epsilon_{eff}(k)=-2J\mathcal{J}_0(A_0)\cos(k).
\end{align}
As a result, the periodic external field effectively rescales the hopping parameter according to $J_{eff}=J\mathcal{J}_0(A_0)$. In particular, whenever $\mathcal{J}_0(A_0)=0$ is satisfied, the quasi-particle dispersion collapses into a flat band and the spreading of initially localized wave packets is suppressed, despite the absence of disorder.

\section{Method}
\label{sec:Method}

In the following, we will focus our attention on periodically driven one-dimensional systems. We note, however, that our formulation admits generalizations to other types of out-of-equilibrium dynamics \cite{Krissia2019, Krissia2020RIXS, Krissia2023core-hole} and higher dimensional systems \cite{Inbar2023,Birkbeck2025,Xiao2026}. In order to compute the spectrum with momentum resolution, as the system is driven away from equilibrium, we employ the extended tunneling spectroscopy protocol introduced in Ref.~\onlinecite{Krissia2019}. The method consists of weakly coupling the desired system to be studied, whose Hamiltonian is denoted as $H_{\mathcal{S}}(t)$, to an auxiliary one-dimensional non-interacting probe chain described by 

\begin{align}
    H_{\mathcal{P}}&=V_g\sum_{i=1}^Ld_i^\dagger d_i,
    \hspace{2mm}
    H_{\mathcal{T}}(t)=g(t)\sum_{i=1}^L\big(c_i^\dagger d_i+h.c.\big),
    \\ \label{composite}
    H(t)&=H_{\mathcal{S}}(t)+H_{\mathcal{T}}(t)+H_{\mathcal{P}}
\end{align}where $c_i^\dagger (c_i)$ and $d_i^\dagger(d_i)$ are the creation (annihilation) operators of the sample and probe, respectively. 

From this point onward, we set the probing pulse as $g(t)=g_0e^{-(t-t_\mathrm{p})^2/2\sigma^2}/\sqrt{2\pi\sigma^2}$, such that the sample and probe are only coupled through a finite time interval. As a result, two advantages over the $g_{\text{step}}(t)= g_0\Theta(t-t_\mathrm{p})$ used in Ref.~\onlinecite{Krissia2019} are achieved. First, as indicated in Fig. \ref{fig:theory}, the instantaneous spectrum will in general be some function of time, but the coupling $g_{\text{step}}(t)$ implicitly assumes $H_{\mathcal{S}}(t)$ is \textit{frozen} during a measurement. This assumption is violated in periodically driven systems. Secondly, and most importantly, the abrupt turn-on of the probe necessarily introduces spectral leakage over a broad energy range, which in our simulations strongly interfered with the Floquet replicas and in many situations led to inconsistent results. The Gaussian coupling avoids both of these problems.

\begin{figure}
    \centering
    \includegraphics[width=1.0\linewidth]{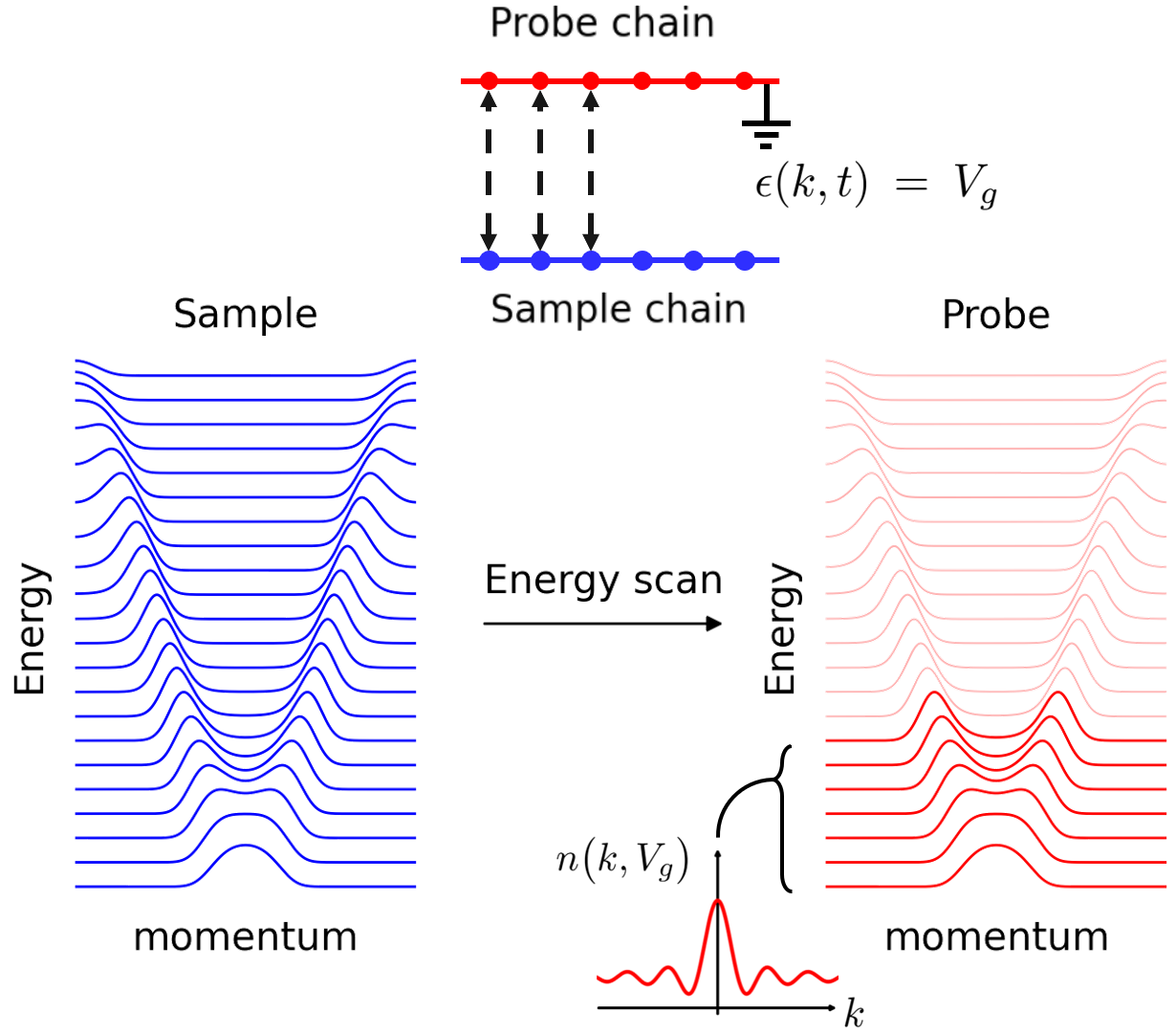}
    \caption{A sample subject to a periodic drive is weakly coupled to a non-interacting probe held at bias voltage $V_g$, where the tunneling amplitude follows a Gaussian envelope. The instantaneous energy spectrum can then be measured by shifting the center of the Gaussian envelope, and computing the momentum distribution of the probe at each time step.}
    \label{fig:sample and probe}
\end{figure}

With this setup at hand, we prepare the sample at some generic state of interest (with the system driven away from equilibrium or not), and set the probe to be empty, that is $|\Psi(t_0)\rangle=|\psi(t_0)\rangle\otimes |0\rangle$. During the probing pulse, the particles are allowed to tunnel from the system to the probe, and the momentum distribution function is measured on the probe chain/detector. In our case, the time evolution is carried out using exact diagonalization (ED) and time-dependent Density Matrix Renormalization Group (tDMRG) \cite{white1992DMRG, white1993DMRG, schollwock2005DMRG, schollwock2011DMRG, vietri, Paeckel2019} methods, but this protocol is generic and can be adapted to any geometry, and any time-dependent solver.

\begin{figure*}
    \centering
    \includegraphics[width=1\linewidth]{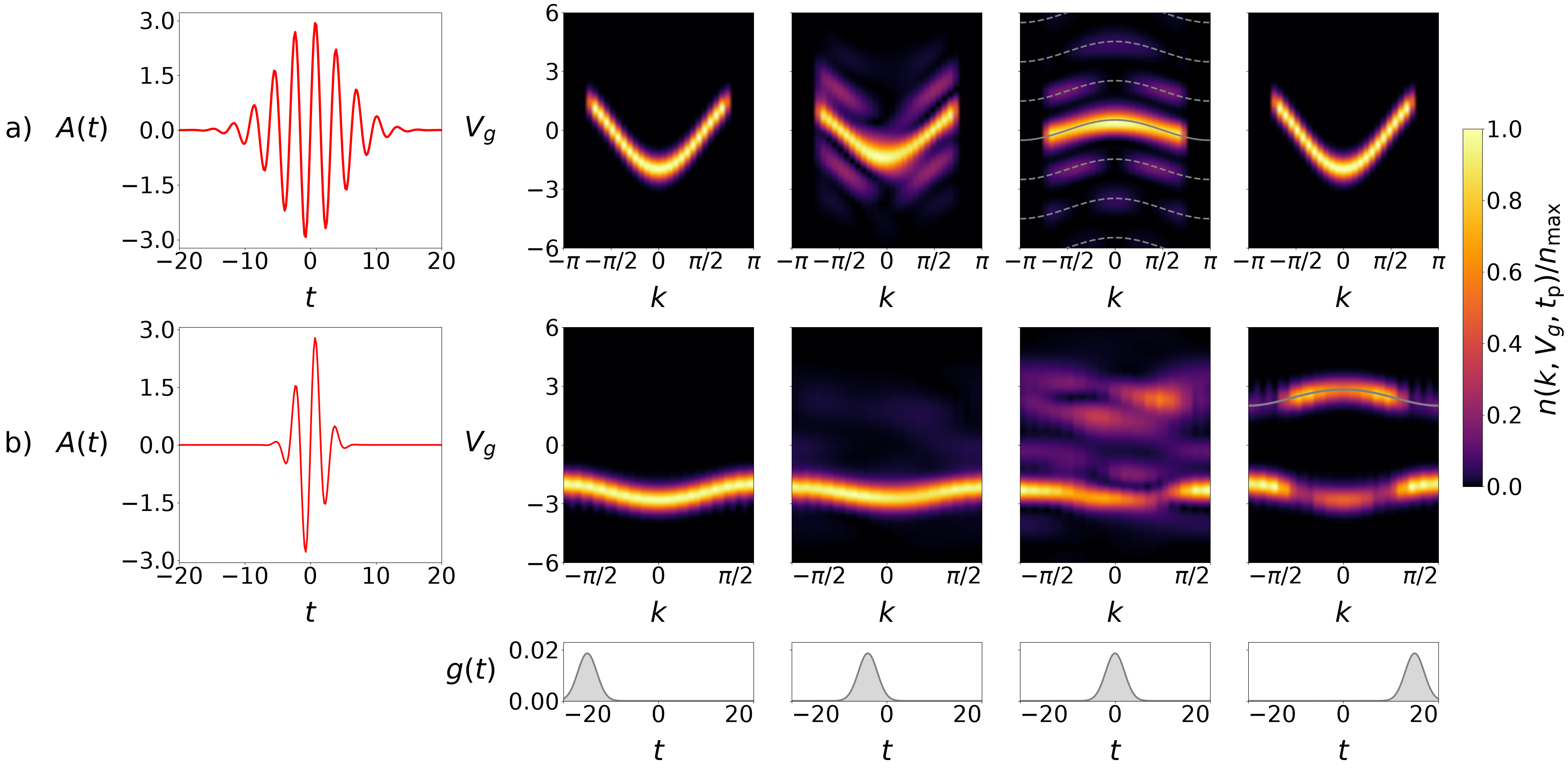}
    \caption{Out-of-equilibrium energy spectrum of a non-interacting fermionic chain driven by $A(t)=A_0\exp(-t^2/2\delta^2)\sin(\Omega t)$. The color scales show the occupation in the probe at the end of the simulation. Each panel is normalized to its own maximum and  $t_0=-20$ is used. Solid and dashed gray lines indicate the instantaneous energy band and its Floquet replicas, respectively. a) Tight-binding model $\Delta=0$ at three-quarter filling, with $(A_0, \Omega, \delta)=(3, 2, 5)$. b) Ionic model $\Delta=2$ at half-filling, with $(A_0, \Omega, \delta)=(3, 2, 2)$.}
    \label{fig:ED}
\end{figure*}

To gain intuition and establish a connection with known Green's functions results, we assume $H_{\mathcal{T}}(t)$ can be treated perturbatively, as discussed in Ref.~\onlinecite{freericks2009positivity}.   The unitary time evolution of the composite system is given by 
\begin{align}
    U_0(t,t_0)&=e^{-iH_{\mathcal{P}}(t-t_0)}\mathcal{T}_t\exp\bigg(-i\int_{t_0}^tH_\mathcal{S}(\tau)d\tau\bigg)
    \\
    U(t, t_0)&\simeq U_0(t, t_0)-i\!\int_{t_0}^t \! U_0(t, \tau)H_\mathcal{T}(\tau)U_0(\tau, t_0)d\tau,
\end{align}
where $\mathcal{T}_t$ denotes time-ordering. After some simple algebraic manipulations we conclude that the momentum distribution of the probe is directly related to the lesser Green's function $G_{ij}^<(\tau,\tau^\prime)=i\langle c_i^\dagger(\tau)c_j(\tau^\prime)\rangle$ of the sample
\begin{align}\label{n plot}
    n(k, \!V_g,\! t_\mathrm{p})&=\frac{1}{L}\sum_{i,j}e^{ik(i-j)}\langle d_i^\dagger(t) d_{j}(t)\rangle
    \\ \label{convolution of Green function}
    \langle d_i^\dagger(t)d_j(t)\rangle&=-i\!\int_{t_0}^t \!\!d\tau \!\!\int_{t_0}^t \!\!d \tau^\prime g(\tau)g(\tau^\prime)e^{iV_g(\tau^\prime-\tau)}G_{ij}^<(\tau,\tau^\prime).
\end{align}
This is precisely the quantity we measure numerically in our simulations: Instead of calculating a two-time correlator and integrating in time, we directly measure an equal time correlator on the probe, a significant simplification with a minimal numerical cost.
Notice that, if instead the probe is initially filled, we can obtain $G_{ij}^>(\tau, \tau^\prime)=-i\langle c_j(\tau^\prime)c_i^\dagger(\tau)\rangle$ by computing the momentum distribution of holes \cite{Cohen2014}. However, since we are primarily interested in determining how excited states are populated by driving the system, we will focus on the first scenario.

Interestingly, for $t-t_\mathrm{p}\gg \sigma$ and $t_\mathrm{p}-t_0\gg \sigma$ the integrals above can be evaluated explicitly by extending their limits to $\pm \infty$. Thus, by decomposing the initial state in a Floquet basis $|\psi(t_0)\rangle=\sum_{\alpha, m} a_\alpha e^{-im\Omega t_0} |\Phi_\alpha^{(m)}\rangle$, we find
\begin{align}\label{n(k, V, t)}
    \frac{n(k,\! V_g, \!t_\mathrm{p})}{g_0^2}\!&=
    \!
    \sum_\beta 
    \!
    \left|\sum_{\alpha, m}a_\alpha e^{i\varepsilon_{\alpha}t_0}e^{i\Delta_{\alpha\beta}^{(m)}t_\mathrm{p}} M_{\alpha\beta}^{(m)}e^{-\sigma^2\Delta_{\alpha\beta}^{(m)}\big.^2/2}\right|^2,
\end{align}
where we defined $\Delta_{\alpha \beta}^{(m)}=V_g-\varepsilon_{\alpha}+\varepsilon_\beta-m\Omega$ and the matrix elements $M_{\alpha\beta}^{(m)}=\sum_n\langle\Phi^{(n)}_\beta|c_k|\Phi^{(n+m)}_\alpha\rangle$. As a result, for a fixed pair $(\alpha, \beta)$ the contributions between distinct replicas $m\neq m^\prime$ are suppressed by $\sim e^{-\sigma^2\Omega^2(m-m^\prime)^2/4}$. Consequently, the parameter $\sigma\Omega=2\pi \sigma /T$ will essentially control the competition between the probe's duration and the period of the drive. Hence, for $\sigma \Omega\gg 1$, only the diagonal terms in (\ref{n(k, V, t)}) contribute to the measured signal and the quasi-energy spectrum is resolved by the condition imposed by the Gaussian peaks, i.e. $V_g=\varepsilon_\alpha-\varepsilon_\beta+m\Omega$. In the opposite limit $\sigma\Omega\ll1$, the sample Hamiltonian can be treated as static, i.e. $H_\mathcal{S}(t_\mathrm{p})$, and the resulting signal follows the instantaneous spectrum. This interplay of time scales can be understood as a manifestation of the \textit{time-energy uncertainty principle}. Namely, the shorter the probing time, the lower the spectral resolution in energy will be.

We note that Eq.~(\ref{convolution of Green function}) has precisely the structure of the time-resolved photoemission signal derived in Ref.~\onlinecite{freericks2009positivity}, with the probing pulse $g(t)$ playing the role of their $s(t)$. The measured signal (i.e., the particle occupation at the probe) is therefore non-negative by construction. This contrasts with the common practice of Fourier transforming the Green function with respect to the relative time at fixed average time, which away from equilibrium does not guarantee a positive spectral density \cite{uhrig2019positivity}. In our numerical approach, positivity is not imposed but follows directly from the fact that the signal is the occupation of a physical probe.

In Appendix \ref{sec:AppendixA} we present a detailed discussion on the application of this spectroscopy protocol for the case of the tight-binding model, where the relevant quantities can be computed analytically.

\begin{figure*}
    \centering
    \includegraphics[width=1\linewidth]{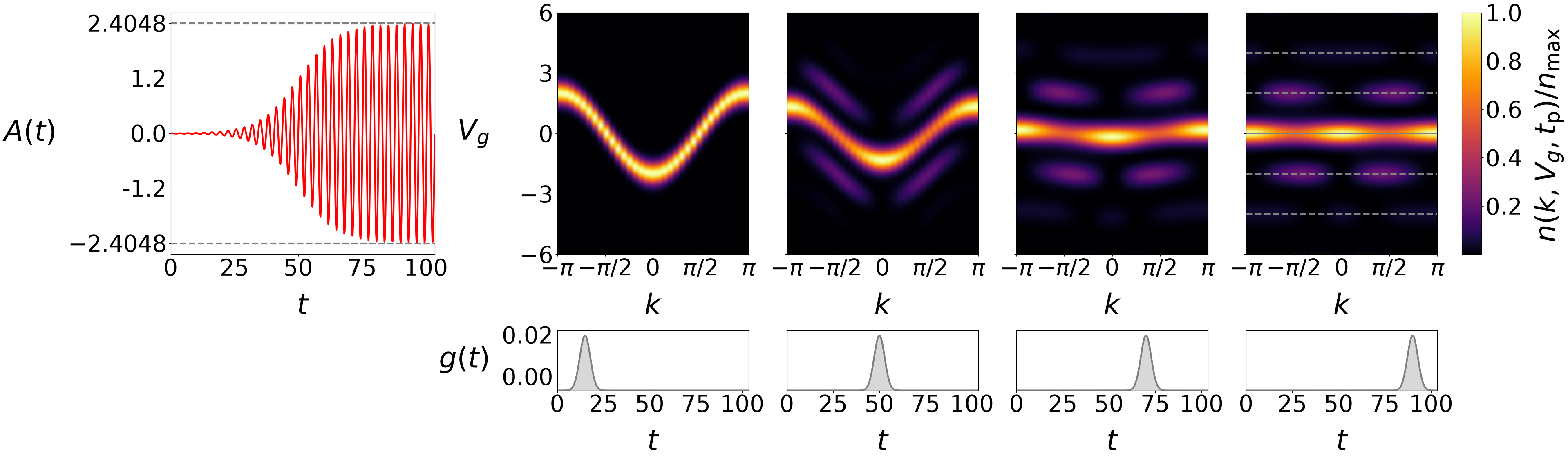}
    \caption{Out-of-equilibrium energy spectrum of the fully filled tight-binding model $\Delta=0$ as it transitions to a dynamically localized state. The drive follows a hyperbolic pulse $A(t)=A_0/2\left[1+\tanh\left((t-t_0)/\delta\right)\right]\sin(\Omega t)$ with $(A_0,\Omega,\delta)=(2.40482\ldots, 2, 15)$, such that the amplitude is tuned to the first zero of $\mathcal{J}_0(A_0)$. The color scales show the occupation in the probe at the end of the simulation. Each panel is normalized to its own maximum and $t_0=0$ is used. Solid and dashed gray lines indicate the instantaneous energy band and its Floquet replicas, respectively.
    }
    \label{fig:dynamical}
\end{figure*}

\section{Results}
\label{sec:Results}

From now on, we consider $L=32$ and $g_0=0.14$, such that the composite system (sample plus probe) remains in the perturbative regime. We also note that each probing time $t_\mathrm{p}$ corresponds to an independent simulation, and that the following snapshots of the spectrum are constructed from Eq.~(\ref{n plot}) evaluated at the end of the simulation. We also adopt $J$ as the unit of energy and $1/J$ as our unit of time.  Moreover, to avoid the appearance of aliasing effects in the time domain, the sampling frequency is selected such that the Nyquist condition, {\it i.e.} $f_{s}>2f_{max}=\Omega/\pi$, is satisfied, while maintaining a sufficient margin above the minimum required value. Finally, as mentioned previously, a tradeoff between the instantaneous and effective energy spectrum is set by the parameter $\sigma \Omega$. Using full width at half maximum (FWHM) as a criterion of resolution, we find $\sigma=2\sqrt{2}$ to be a reasonable intermediate choice for the frequency range considered by us. This corresponds to a temporal resolution of $\delta t=2\sqrt{\ln 2}\sigma\approx 4.71$ and an energy resolution of $\delta E=2\sqrt{\ln 2}/\sigma\approx 0.59$. As a result, we are able to probe the sub-cycle regime for $\Omega \lesssim 1.5$, while still separating different spectral features. We emphasize, however, that this choice is not unique.

We begin by presenting results using ED for the non-interacting case, to both benchmark the technique and showcase how much physics can already be captured at the system sizes accessible to ED. Next, we examine both spinless and spinful systems in their strongly interacting limits. In those cases, we utilize the TeNPy library \cite{tenpy2024} to carry out the tDMRG \cite{white1992DMRG, white1993DMRG, schollwock2005DMRG, schollwock2011DMRG, vietri, Paeckel2019} calculations. This setup enabled us to perform 200 simulations in parallel, with bond dimension $\chi=300$, to scan over the different probing energies, while maintaining the maximum truncation error below $10^{-5}$. 

\subsection{Exact Diagonalization: non-interacting models}
\label{sec:Exact Diagonalization}

As a first application, we consider a periodically driven non-interacting fermionic chain in the presence of a staggered ionic potential
\begin{align}\label{quadratic model}
    H&=-J\sum_j \bigg(e^{-iA(t)}c^\dagger_jc_{j+1}+h.c.\bigg)
    +
    \Delta\sum_j (-1)^j n_j
\end{align}
where again $J$ denotes the bare hopping parameter, $A(t)$ the driving field and $(-1)^j\Delta$ the alternating on-site potential. Since the Peierls phase introduces a shift in momentum, {\it i.e.} $k\rightarrow k(t)=k-A(t)$, the instantaneous band structure becomes 
\begin{align}
    \epsilon^{\pm}(k, t)=\pm \sqrt{\Delta^2+ \bigg(2J\cos\big(k-A(t)\big)\bigg)^2}.
\end{align}

Therefore, this model provides an interesting test case to benchmark our method, allowing one to explore several key non-equilibrium phenomena, such as: (i) Formation of Floquet replicas in the sub-cycle regime, (ii) generation of long-lived excitations across the ionic gap, and (iii) the transition to dynamically localized states. With this purpose in mind, we couple (\ref{quadratic model}) to a probe chain by following the prescription discussed in Sec. \ref{sec:Method}. 

\begin{figure*}
    \centering
    \includegraphics[width=0.85\linewidth]{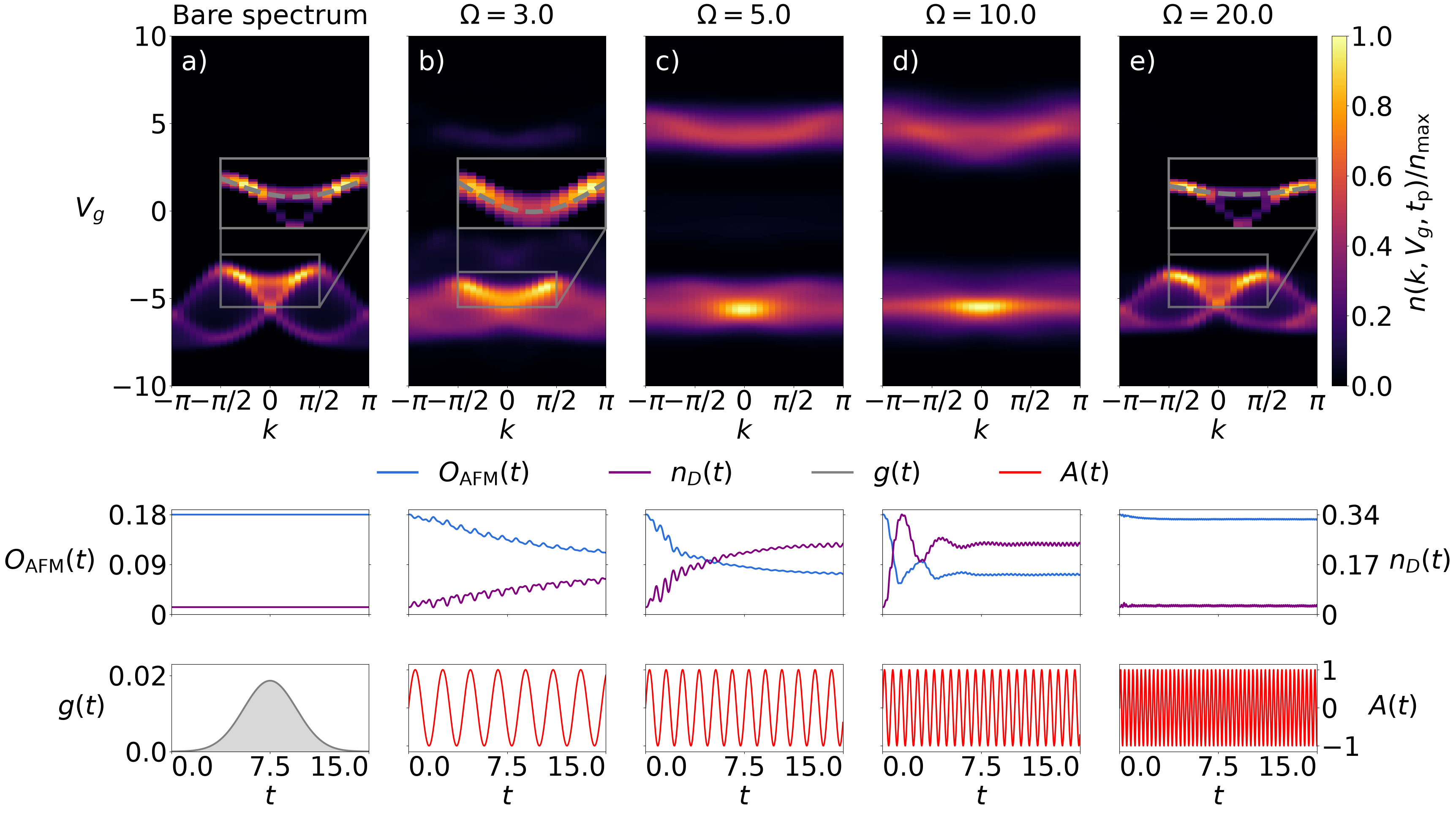}
    \caption{Out-of-equilibrium energy spectrum of the half-filled Hubbard model at different driving frequencies. Top row: Probe occupation $n(k, V_g, t_\mathrm{p})$ evaluated at time $t_\mathrm{max}=15.0$ for (a) the undriven system (bare spectrum) and (b)-(e) increasing driving frequencies $\Omega$. Each panel and inset is normalized to its own maximum. Insets in (a), (b) and (e) show the second derivative $-\partial^2 n(k, V_g, t_\mathrm{p})/\partial V_g^2$, together with the theoretical spinon dispersion of Eq.~(\ref{J_S}) (dashed gray) for comparison. See Fig.~\ref{fig:spinon} for more details. Middle row: Staggered magnetization $O_\mathrm{AFM}(t)$ (blue, left axis) and doublon density $n_D(t)$ (purple, right axis) as functions of time. Bottom row: Probe pulse $g(t)$ (gray, left axis), identical for all panels and centered at $t_\mathrm{p} = 7.5$, and driving field $A(t) = \Theta(t) A_0 \sin(\Omega t)$ (red, right axis) with $A_0 = 1.0$.}
    \label{fig:hubbard}
\end{figure*}

The first two scenarios are illustrated in Fig. \ref{fig:ED}a and Fig. \ref{fig:ED}b, respectively, where a Gaussian envelope containing few cycles per pulse drives the system. To this end we set $\Delta=0$ in Fig. \ref{fig:ED}a and $\Delta=2$ in Fig. \ref{fig:ED}b. The goal here is to identify which spectral features emerge under minimal external perturbation, for both gapped and gapless states. In each case, the system is initially prepared at its ground state before pumping, and independent simulations are performed at different probing times $t_{\mathrm{p}}$. In the gapless system, we observe at the center of the driving pulse the emergence of Floquet replicas, as well as their transient regime at intermediate times. Notably, in Fig. \ref{fig:ED}a the spectrum becomes inverted as the drive achieves its maximum, since for the given amplitude we have $\mathcal{J}_0(A_0)<0$ \cite{weitenberg2021tailoring, tsuji2011dynamical}. We further observe that successive replicas alternate between bright and dark along their bands, an effect we trace in Appendix \ref{sec:AppendixA} to a momentum-dependent selection rule set by the parity of the replica index.

Alternatively, by setting the gapped system at half-filling, such that only the lower band is populated, we see that although the driving frequency is non-resonant, {\it i.e.} $2\Delta\neq \Omega$, multi-photon processes still dominate and induce inter-band transitions. However, once the drive is removed and, in the absence of other dissipation mechanisms, the system is trapped in this excited configuration. Such an effect can be interpreted as heating, once the system has effectively absorbed some of the energy introduced by the drive. 

Finally, the transition to a dynamically localized state can be probed by gradually increasing the driving amplitude until $\mathcal{J}_0(A_0)=0$ is satisfied. We set $\Delta=0$ for simplicity in this case. As shown in Fig. \ref{fig:dynamical}, besides the formation of Floquet replicas, the energy band progressively collapses into a flat dispersion at late times, indicating that the system has reached a steady state in which transport is suppressed. However, we notice that in the process of preparation of all three systems we have populated higher Floquet bands and, as a result, they cannot be assumed to be described by the effective Floquet Hamiltonian (\ref{effective Hamiltonian}). Understanding how to avoid this type of heating during the onset of the drive is an active research area \cite{Anatoli2021fermi-golden-rule, cupo2025Dartmounth}, and beyond the scope of this work.

\subsection{DMRG: Interacting models}
\label{sec:DMRG}

We now turn to interacting systems to investigate the robustness of our technique in the presence of strong correlations and to assess its capability to probe regimes beyond the high-frequency expansion.

\subsubsection{Spinful fermions: Hubbard model}
\label{sec:spinful}

\begin{figure}
    \centering
    \includegraphics[width=0.95\linewidth]{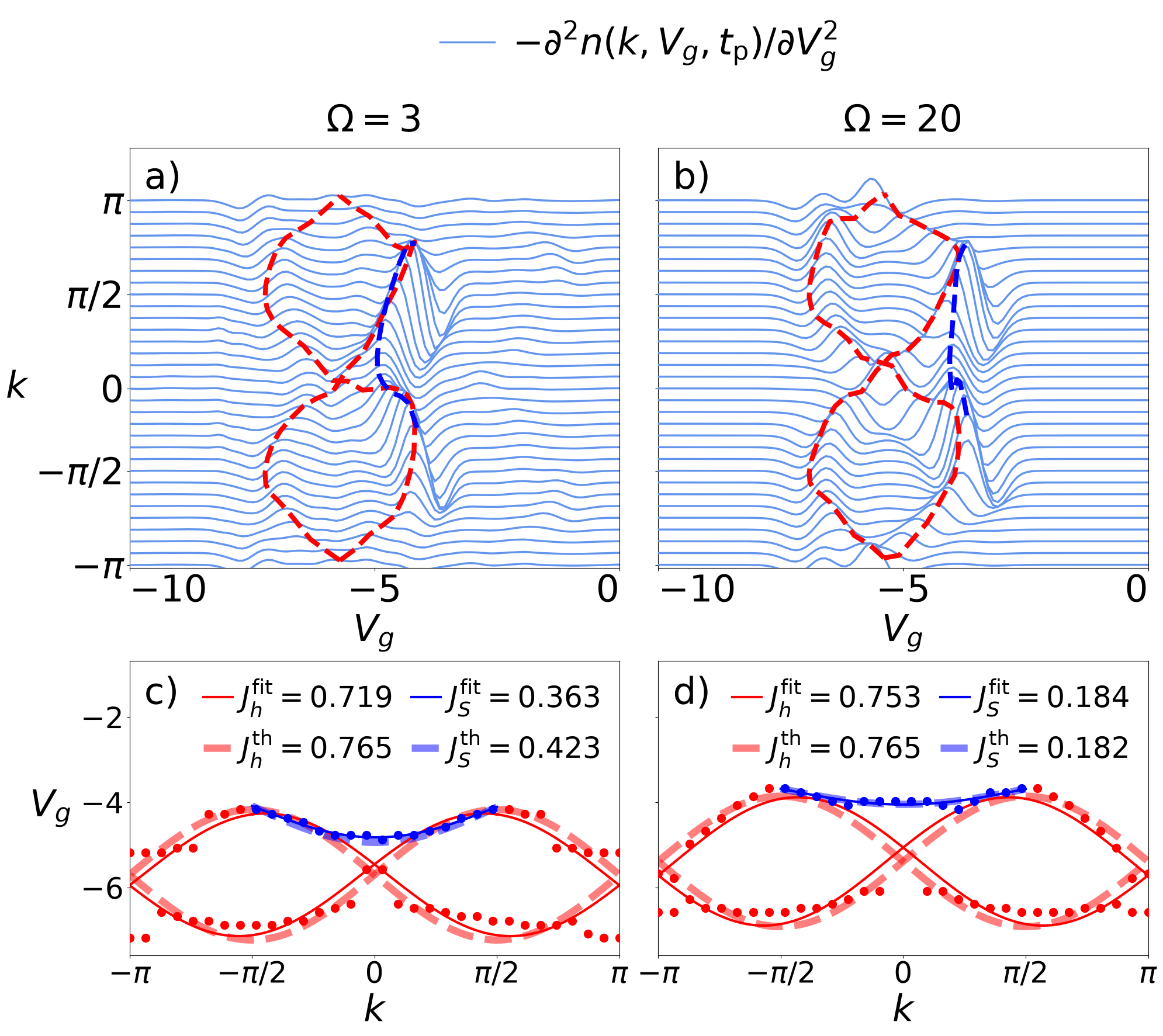}
    \caption{Second derivative of the probe occupation $-\partial^2 n(k, V_g, t_\mathrm{p})/\partial V_g^2$ at (a) $\Omega = 3$ and (b) $\Omega = 20$, computed from the spectra in Fig.~5. Curves are offset vertically according to momentum, and each panel is normalized to its own maximum. Panels (c) and (d) show the peak positions of the spinon (blue) and the two holon bands (red), which are fitted by $V_g(k) = -2J^{\mathrm{fit}}\cos(k + \delta k) + c$ (solid lines). Dashed lines are plotted with  $J^{\mathrm{fit}}=J_h^\mathrm{th}=J\mathcal{J}_0(A_0)$ and $J^{\mathrm{fit}}=J_S^\mathrm{th}$ as in Eq.~(\ref{J_S}), for comparison with the theoretical prediction.}
    \label{fig:spinon}
\end{figure}

To explore the competition between strong interactions and periodic driving, we consider the Hubbard model
\begin{align}\nonumber
    H(t)=-J\sum_{j,\sigma} \bigg(e^{-iA(t)}c^\dagger_{j\sigma}c_{j+1\sigma}+h.c.\bigg)&
    \\ \label{hubbard}
    +U\sum_j \left(n_{j\uparrow}-\frac{1}{2}\right)\left(n_{j\downarrow}-\frac{1}{2}\right),&
\end{align}
where $c^\dagger_{j\sigma}$ creates an electron with spin $\sigma\in\{\uparrow, \downarrow\}$ and $c_{j\sigma}$ annihilates it, while $U$ parametrizes the on-site interaction and $A(t)$ the driving field.

As the system is driven away from equilibrium, a more natural description of (\ref{hubbard}) is obtained by moving to a rotating frame. Following the approach of \cite{Anatoli2016Schrieffer-Wolff, kumar2021inducing}, we define the time-dependent unitary operator
\begin{align}
    \mathcal{R}(t)&=\exp\left[-it U\sum_j \left(n_{j\uparrow}-\frac{1}{2}\right)\left(n_{j\downarrow}-\frac{1}{2}\right)\right],
\end{align}
such that the evolution of the state $|\psi_\mathcal{R}(t)\rangle=\mathcal{R}^\dagger(t)|\psi(t)\rangle$ is now dictated by $i\partial_t|\psi_\mathcal{R}(t)\rangle= H_\mathcal{R}(t)|\psi_\mathcal{R}(t)\rangle$. By using the expression $e^{i\alpha n_{j\sigma}}=(1-n_{j\sigma})+e^{i\alpha}n_{j\sigma}$ and (\ref{Jacobi-Anger}), we find
\begin{align}\nonumber
    H_\mathcal{R}(t)&=-J\!\sum_{j\sigma m}\sum_{\nu\nu^\prime}
    \mathcal{J}_m(A_0)\bigg(e^{it\Delta_{\nu\nu^\prime}^{(m)}}c^{(\nu)\dagger}_{j\sigma}c^{(\nu^\prime)}_{j+1\sigma}+h.c.\bigg),
    \\
    c_{j\sigma}^{(0)}&=(1-n_{j\bar\sigma})c_{j\sigma},
    \hspace{5mm}
    c_{j\sigma}^{(1)}=n_{j\bar\sigma}c_{j\sigma},
\end{align}
where $\bar\sigma$ denotes a spin flip, i.e. $\{\bar\uparrow,\bar\downarrow\}=\{\downarrow, \uparrow\}$, and $\Delta_{\nu\nu^\prime}^{(m)}=(\nu-\nu^\prime)U-m \Omega$. Here $\nu\in\{0, 1\}$ is the number of $\bar\sigma$ electrons on site $j$, such that $c_{j\sigma}^{(0)}$ creates a holon and $c_{j\sigma}^{(1)}$ destroys a doublon. 
To further ensure the periodicity of the driven system, we consider $U$ and $\Omega$ to be commensurate with each other, that is, $U/\Omega=p/q$, for $p$ and $q$ co-prime integers. 
As a result, each tunneling process becomes dependent on the energy difference of creating or destroying a doublon-holon pair, for which $\nu\neq\nu^\prime$. The dynamics is thus separated into resonant and off-resonant regimes, depending if these energy differences match or not a multiple of the driving frequency
\begin{align}\label{resonant condition}
    (\nu-\nu^\prime)U=m\Omega.
\end{align}
Since this condition is trivially satisfied for $m=0$ and $\nu=\nu^\prime$, the usual $\mathcal{J}_0(A_0)$ renormalization is always present in the time-averaged Hamiltonian. In contrast, if (\ref{resonant condition}) holds for a given non-zero $m$, the corresponding tunneling channel is no longer suppressed by fast oscillations and provides an additional $\mathcal{J}_m(A_0)$ contribution to the effective Hamiltonian. Resonantly driving the system thus favors the formation of real excitations.

Interestingly, at half-filling and strong interactions $J\ll U$ the low energy dynamics is confined to the manifold of singly occupied states. Consequently, since a tunneling event necessarily creates a doublon-holon pair, the zeroth order in the high-frequency expansion vanishes. The next non-vanishing contribution arises at linear order, where virtual states outside this subspace mediate spin interactions. Thus, upon using the projector $\mathcal{P}=\prod_j (1-n_{j\uparrow}n_{j\downarrow})$ and considering the Fourier expansion $H_\mathcal{R}(t)=\sum_m e^{im\Omega_0 t}H_\mathcal{R}^{(m)}$, where $\Omega_0=U/p=\Omega/q$ denotes fundamental frequency of the system,  we find
\begin{align}
    H_{eff}&
    =
    \mathcal{P}\!
    \sum_{m=1}^\infty
    \frac{[H^{(m)}_\mathcal{R},H^{(-m)}_\mathcal{R}]}{m\Omega_0}\mathcal{P}
    =
    J_S\sum_jS_j\cdot S_{j+1}, 
    \\ \label{J_S}
    J_S&=4J^2\sum_m\frac{\mathcal{J}_m(A_0)^2}{U-m \Omega}.
\end{align}
    For $J \ll U \ll \Omega$, we have $J_S \approx 4J^2\mathcal{J}_0(A_0)^2/U$, whereas in the opposite limit, $J \ll \Omega \ll U$, the spin exchange is unaffected by the drive $J_S \approx 4J^2/U$. 

\begin{figure*}
    \centering
    \includegraphics[width=1.0\linewidth]{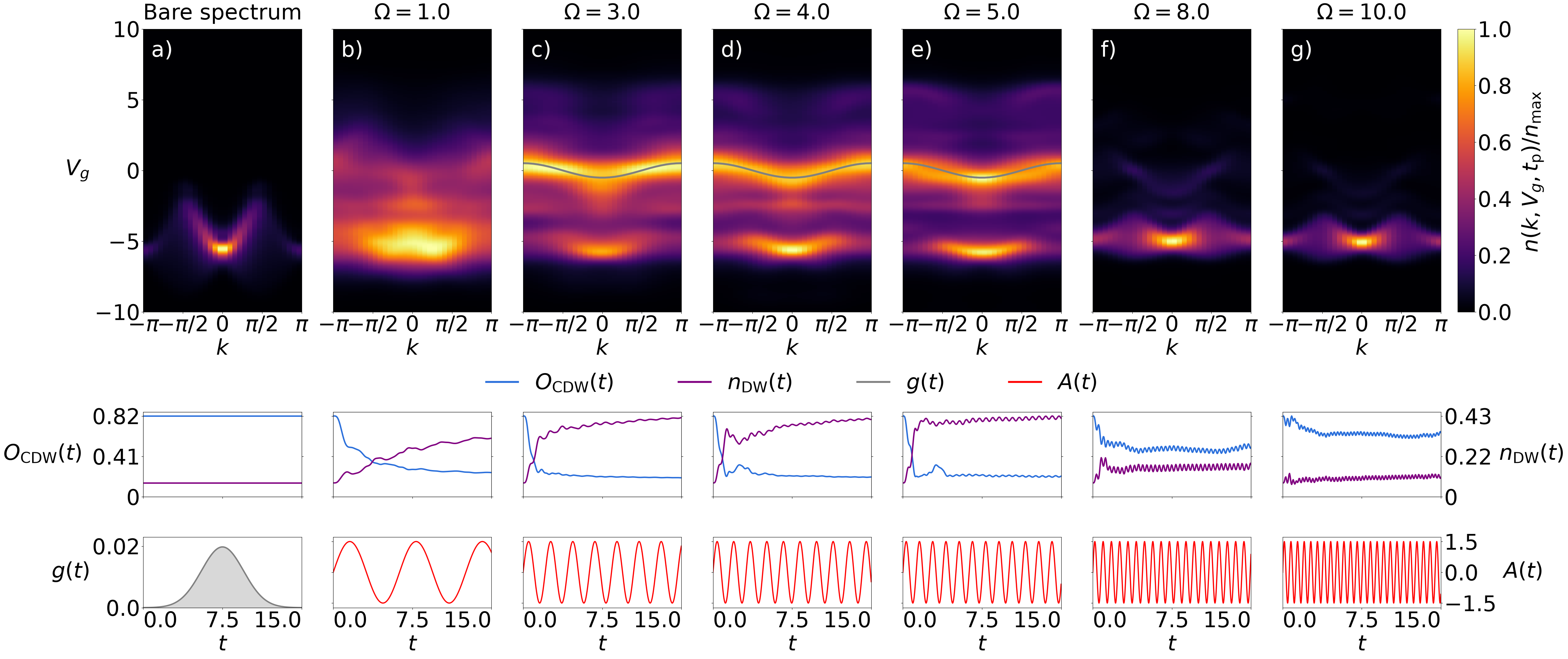}
    \caption{Out-of-equilibrium energy spectrum of interacting spinless fermions at half-filling for different driving frequencies. Top row: Probe occupation $n(k, V_g, t_\mathrm{p})$ evaluated at $t_\mathrm{max}=15.0$ for (a) the undriven system (bare spectrum) and (b)-(g) increasing driving frequencies $\Omega$. Each panel is normalized to its own maximum. Middle row: Charge density wave order parameter $O_\mathrm{CDW}(t)$ (blue, left axis) and domain-wall density $n_\mathrm{DW}(t)$ (purple, right axis) as functions of time. Bottom row: Probe pulse $g(t)$ (gray, left axis), identical for all panels and centered at $t_\mathrm{p} = 7.5$, and driving field $A(t) = \Theta(t) A_0 \sin(\Omega t)$ (red, right axis) with $A_0 = 1.5$.}
    \label{fig:tV}
\end{figure*}

In Fig.~\ref{fig:hubbard} we compare these predictions to the measured spectra of our proposed spectroscopy protocol. To this end, we prepare the system in the half-filled ground state with $U=10$. The drive is taken to be $A(t)=\Theta(t)A_0\sin(\Omega t)$ and the probing pulse is centered at $t_{\mathrm{p}}=t_{\mathrm{max}}/2=7.5$. As expected, the equilibrium spectrum reproduces the characteristic charge and spin branches of spin-charge separation \cite{2005Hubbard-book, Haldane1981luttinger, giamarchi2004quantum}. 
Moreover, at resonant driving, i.e. $\Omega=5$ and $\Omega=10$, the Lower Hubbard Band (LHB) loses most of its structure, while transferring a considerable amount of spectral weight to the Upper Hubbard Band (UHB). As a result, we should expect a reduction in antiferromagnetic correlations and an increase in the density of doubly occupied sites. We confirm this by also computing the staggered magnetization and the density of doublons 
\begin{align}\label{AFM}
    O_{AFM}^{\hspace{1mm}2}&=\frac{1}{L^2}\sum_{ij} (-1)^{i+j}
        \langle S^z_i S^z_j \rangle,
    \\ \label{doublons}
    n_D&=\frac{1}{L}\sum_j\langle n_{j\uparrow}n_{j\downarrow}\rangle.
\end{align}
The time evolution of both quantities is shown in the middle panels of Fig.~\ref{fig:hubbard}.

Away from resonance, the holon and spinon branches remain well defined and can be further resolved by taking the second derivative of the measured signal, i.e. $-\partial^2 n(k, V_g, t_\mathrm{p})/\partial V_g^2$, which we highlight in the insets of panels (a), (b) and (e) of Fig.~\ref{fig:hubbard}. Finally, by fitting the peak positions (see Fig.~\ref{fig:spinon}), we find that the extracted values reasonably agree with the theoretical predictions set by $J_h=J\mathcal{J}_0(A_0)$ and $J_S$ as in Eq.~(\ref{J_S}), which is only renormalized by the drive in the limit $J\ll U\ll \Omega$.

\subsubsection{Spinless fermions: t-V chain}
\label{sec:spinless}

We also consider a periodically driven chain of interacting spinless fermions
\begin{align}\nonumber
    H=-J\sum_{j} \bigg(e^{-iA(t)}c^\dagger_{j}c_{j+1}+h.c.\bigg)&
    \\ \label{NN}
    +V\sum_j \left(n_{j}-\frac{1}{2}\right)\left(n_{j+1}-\frac{1}{2}\right)&,
\end{align}
where $c_j^\dagger$ ($c_j$) denotes the creation (annihilation) operators, $V$ parametrizes the nearest-neighbor interaction and $A(t)$ the driving field. At half-filling and strong interactions, {\it i.e.} $J\ll V$, the resulting ground state exhibits a two-fold degeneracy characterized by the alternating patterns $|0101...\rangle$ and $|1010...\rangle$. As a result, translational symmetry is spontaneously broken and the unit cell doubles in size. In this regime, the low-energy excitations are described by domain wall defects connecting these two different charge orders.

As first noticed in Ref.~\onlinecite{Germans2023InGap}, this system exhibits the formation of an in-gap dispersion when driven close to resonance, i.e. $V\simeq\Omega$. Their numerics further showed that strong interactions are essential for the emergence of this band, and that its bandwidth is independent of $V$. The authors therefore conjectured that such an effect probably originates from ``Villain-like modes'', i.e. propagating domain walls. However, as they point out, a microscopic description of this band is still missing. For completeness, in the following we provide such a description, which further supports their claim.

To explain this effect we define the time-dependent unitary operator 
\begin{align}\label{R for tV chain}
    \mathcal{R}(t)=\exp\left[it V \sum_j \left(n_j-\frac{1}{2}\right)\left(n_{j+1}-\frac{1}{2}\right)\right],
\end{align}
and use the identity $e^{i\alpha n_{j}}=(1-n_{j})+e^{i\alpha}n_{j}$ to express the Hamiltonian in the rotating frame as
\begin{align} \label{t-V chain rotating}
    H_{\mathcal{R}}(t)&=-J\sum_m\Bigg[\mathcal{J}_m(A_0)e^{-im\Omega t}\times
    \\ \nonumber
    &\hspace{1cm}\times\bigg(T_0+e^{iVt}T_++e^{-iVt}T_-\bigg)+h.c.\Bigg]
\end{align}
\begin{align}
    T_+&=\sum_j(1-n_{j-1})c_j^\dagger c_{j+1}n_{j+2}
    \\
    T_-&=\sum_jn_{j-1}c_j^\dagger c_{j+1}(1-n_{j+2})
    \\
    T_0&=\sum_j\bigg((1-n_{j-1})c_j^\dagger c_{j+1}(1-n_{j+2}) 
    \\ \nonumber
    & \hspace{3cm} + n_{j-1}c_j^\dagger c_{j+1}n_{j+2}\bigg)
\end{align}
where the operators $T_0$, $T_+$ and $T_-$ describe hopping processes that preserve, create, and annihilate a domain wall, respectively. (More details on this derivation are provided in the Appendix \ref{sec:appendixB}).

Consequently, away from resonance, the time-averaged Hamiltonian is characterized solely by the $m=0$ Fourier component of the $T_0$ channel. On the other hand, at resonance, the fast oscillating phases accompanying the $T_\pm$ channels are compensated, and the creation or annihilation of domain walls is enabled by the absorption or emission of photons, respectively. The resulting defects then propagate under the action of $T_0$, thus giving rise to the dispersion
\begin{align}
\epsilon_{\mathrm{DW}}(k)=-2J\mathcal{J}_0(A_0)\cos(k).
\end{align}


The equilibrium spectrum, by contrast, is determined by adding an electron on top of the charge ordered ground state, \textit{i.e.} $|...011101...\rangle$, which necessarily creates {\it two} domain walls, with an energy cost $\simeq 2V$. 
However, given that the drive induces the formation of particle-conserving excitations, the energetically unstable states $|...011001...\rangle$ become part of the low energy dynamics. Hence, the excitations induced by the drive at resonance have an energy cost equal to $V$, placing the $\epsilon_{DW}(k)$ dispersion in the middle of the equilibrium energy gap.

\begin{figure}
    \centering
    \includegraphics[width=.9\linewidth]{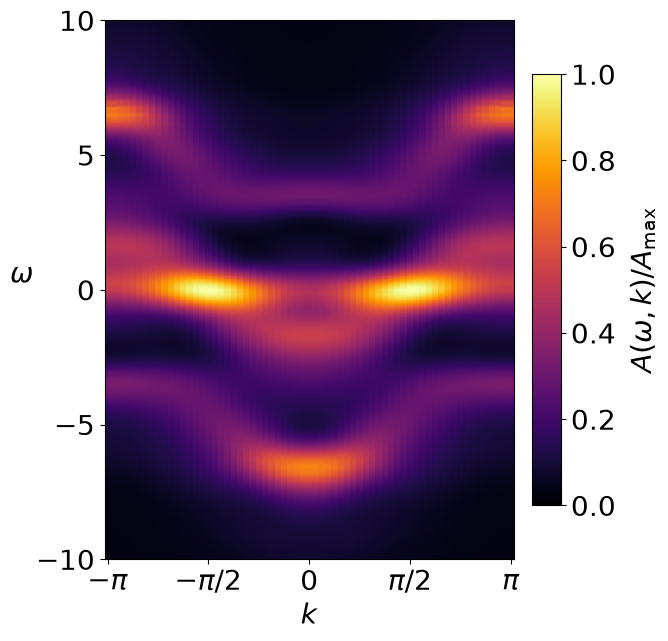}
    \caption{Finite temperature spectrum of the interacting spinless fermionic model with $V=5$ obtained with finite-temperature tDMRG, at inverse temperature $\beta=1/T=2/J$ in the grand canonical ensemble, with chemical potential $\mu=0$. The spectral function is indicated by the color scale, which is normalized to its own maximum.}
    \label{fig:temperature}
\end{figure}

This observation is further confirmed in Fig. \ref{fig:tV}, where we consider $V=5$ and prepare the system at the half-filled ground state. We also take $A(t)=\Theta(t)A_0\sin(\Omega t)$ and set $t_{\mathrm{p}}=t_{\mathrm{max}}/2=7.5$. At resonance $\Omega=5$ and near-resonance, i.e. $3\leq\Omega<5$, an in-gap state is observed, but it becomes absent for other driving frequencies. We also note that the appearance of the in-gap state is always accompanied by a suppression of the CDW order and an increase in the domain-wall density
\begin{align}
    O_{CDW}^{\hspace{1mm}2}&=\!\frac{4}{L^2}\!\sum_{ij}(-1)^{i+j}\bigg\langle\!\!\left(n_{i}-\frac{1}{2}\right)\left(n_{j}-\frac{1}{2}\right)\!\!\bigg\rangle,
    \\
    n_{DW}&=\frac{1}{L-1}\!\sum_{j}\!\bigg(n_{j}n_{j+1}+(1\!-\!n_{j})(1\!-\!n_{j+1})\!\bigg).
\end{align}
As shown in the middle panels of Fig.~\ref{fig:tV}. Our derivation is also consistent with the observation made in Ref.~\onlinecite{Germans2023InGap}, where the bandwidth of this in-gap dispersion was numerically proven to be independent of $V$.

Finally, although the non-interacting ionic model presents a similar staggered charge order, the excitations induced by the drive match the gap of the system in equilibrium, and no in-gap states are expected to emerge in this case. Furthermore, as discussed in Ref.~\onlinecite{nocera2018finite-temperature}, in-gap states also do not appear in the single-particle spectrum of the Hubbard model, which is primarily dominated by spin-charge separation and anomalous spectral weight transfer \cite{Eskes1991}.

\subsubsection{In-gap formation at finite temperature}

The emergence of an in-gap band is not a phenomenon unique to periodically driven systems. As also suggested in Ref.~\onlinecite{Germans2023InGap}, it can be attributed to ``neutral excitations'', which as discussed previously correspond to the creation of single domain walls on top of the charge ordered ground state. This also occurs at finite temperature, since such defects can be realized by thermal excitations. 

This is indeed manifest in the finite-temperature spectrum, Fig. \ref{fig:temperature}, obtained with tDMRG in imaginary time at inverse temperature $\beta=1/T=2/J$. This result can be compared to those obtained by exact diagonalization in Ref.~\onlinecite{Germans2023InGap}. We used the grand canonical ensemble and bond dimension $\chi=400$ in our simulations, that extended to a time window of width $t_{\mathrm{max}}=30$ (see Refs.\cite{feiguin2005a,vietri,Karrasch2013} for technical details).

To gain some insight behind the in-gap band presented in Fig. \ref{fig:temperature}, we consider (\ref{NN}) in the atomic limit, such that the spectral function at finite temperature can be determined analytically for the grand canonical ensemble. By exploiting transfer matrix techniques, see Appendix \ref{AppendixC} for details, we obtain 
\begin{align} \label{spectral function}
    A(\omega)
    \!
    &=
    \!
    \frac{1+z^2}{2(1+z)^2}
    \bigg[
    \delta(\omega+V)
    \!+\!
    \delta(\omega-V)
    \bigg] 
    \!+\!
    \frac{2z}{(1+z)^2}\delta(\omega),
\end{align}
where $z=e^{\beta V/2}$ is the fugacity at half-filling $\mu=V$. Consequently, at zero temperature $T=0\Rightarrow z\rightarrow\infty$ the spectral weight is concentrated entirely in the two atomic peaks at $\omega=\pm V$, while for any finite temperature a third peak develops at $\omega=0$ with weight controlled by $\sim 2e^{-\beta V/2}$. Periodic driving is therefore only one of several ways of creating an in-gap neutral excitation.

\section{Conclusion}
\label{sec:Conclusion}

In this work, we introduced a time and momentum resolved tunneling spectroscopy protocol to investigate the quasi instantaneous time-resolved energy spectrum of periodically driven quantum systems. Given its close connection to tunneling-based experiments, we anticipate our technique to become a powerful tool in the characterization of Floquet driven dynamics.  

Our approach can resolve the emergence of Floquet replicas in real-time and consistently explore their transient regimes during state preparation, as well as other effects such as renormalization of the bandwidth, generation of photon-assisted inter-band transitions and formation of dynamically localized states. 
Furthermore, our protocol can detect the different responses of the charge and spin sectors under periodic driving, and describe the transfer of spectral weight between replicas (heating). 

Finally, in the interacting spinless fermion model, we provided insights on the microscopic mechanism behind the emergent in-gap states, showing that under resonant driving  or in the presence of a thermal bath, the creation of domain walls give rise to additional spectral features.

\begin{acknowledgments}
    We thank K. Zawadzki, M. A. Sentef, A. Shilcusky, P. Weinberg, G. Delfino and E. Habjan for the valuable discussions. We acknowledge the generous computational resources provided by Northeastern University’s Explorer Cluster at the Massachusetts Green High Performance Computing Center (MGHPCC). The authors thank the U.S.~Department of Energy, Office of Basic Energy Sciences, under Award Number DE-SC0022311.
\end{acknowledgments}

\section*{Data Availability}
The data that support the findings of this article, together with the
scripts used to generate and analyze them, are publicly available in the
Zenodo repository \cite{Queiroz-2026} and on GitHub \cite{Queiroz-2026-github},
where animations of each simulation can also be found.

\appendix

\section{Spectroscopy response of periodically driven tight-binding model}
\label{sec:AppendixA}

As a pedagogical example, we explore the response of a periodically driven tight-binding model under our spectroscopy protocol. Given the translational invariance of the system, each momentum sector reduces to a two-level system with well defined momentum $k$ described by 
\begin{align}
    H_k(t)=\left(
        \begin{array}{cc}
            \epsilon(k,t) & g(t) \\
             g(t) & V_g 
        \end{array}
    \right)
\end{align}
where $\epsilon(k,t)=-2J\cos\big(k-A_0\sin\big(\Omega t\big)\big)$ denotes the instantaneous dispersion, $g(t)=g_0e^{-(t-t_\mathrm{p})^2/2\sigma^2}/\sqrt{2\pi\sigma^2}$ the probing pulse and $V_g$ the bias voltage.

To isolate the tunneling dynamics between the sample and the probe, we remove the diagonal terms in the Hamiltonian by introducing the time-dependent unitary operator
\begin{align}
    \mathcal{R}_k(t)=\left(
        \begin{array}{cc}
            e^{i\phi_k(t)} & 0 \\
            0 & e^{-iV_gt}
        \end{array}
    \right)
\end{align}
such that
\begin{align}
    \phi_k(t)=&\int_0^t \epsilon(k,\tau)d\tau
    = \epsilon_{eff}(k)t+\delta \phi_k(t)
    \\
    \delta\phi_k(t)=&\sum_{m\neq0}\delta\phi^{(m)}_k(t)
    \\ \nonumber
    =&\sum_{m\neq0}J\mathcal{J}_m(A_0)\bigg(e^{ik}+(-1)^m e^{-ik}\bigg)\frac{e^{-im\Omega t}-1}{im\Omega},
\end{align}
with $\epsilon_{eff}(k)=-2J\mathcal{J}_0(A_0)\cos(k)$. As a result, the transformed Hamiltonian becomes 
\begin{align}\nonumber  
    H_{k}^{[\mathcal{R}_k]}&=\mathcal{R}_k(t) H_k \mathcal{R}_k^\dagger(t)-i \mathcal{R}_k(t) \partial_t \mathcal{R}_k^\dagger(t)
    \\
    &=
    \left(
        \begin{array}{cc}
            0 & g(t)e^{iV_gt-i\phi_k(t)}
        \\
        g(t)e^{-iV_gt+i\phi_k(t)} & 0
        \end{array}
    \right).
\end{align}

Furthermore, by projecting the wave-function onto the occupation basis,
$|\psi_k(t)\rangle=\alpha_k(t)|k,0\rangle+\beta_k(t)|0,k\rangle$, the Schrödinger equation yields
\begin{align}
    i\dot\alpha_k
    &=
    g(t)e^{iV_gt-i\phi_k(t)}
    \beta_k,
    \\
    i\dot\beta_k
    &=
    g(t)e^{-iV_gt+i\phi_k(t)}
    \alpha_k,
\end{align}
which in the weak coupling regime $g_0\ll 1$ admits a perturbative solution. Assuming the probe is initially empty, {\it i.e.} $\alpha_k(-\infty)=1$ and $\beta_k(-\infty)=0$, we obtain to first order 
\begin{align}
    \beta_k(t)=-i\int_{-\infty}^tg(\tau)e^{-i\big(V_g-\epsilon_{eff}(k)\big)\tau+i\delta\phi_k(\tau)}d\tau.
\end{align}
The momentum distribution of the probe then reads

\begin{align}
    n(k, V_g, t_\mathrm{p})
    =
    g_0^2
    \left|\sum_m \mathcal{F}^{(m)}_ke^{im\Omega t_\mathrm{p}}e^{-\sigma^2\big(V_g-\Delta_k^{(m)}\big)^2/2}\right|^2,
\end{align}    
where we took the limit of long probing times $t\rightarrow\infty$, defined $\Delta_k^{(m)}=\epsilon_{eff}(k)+m\Omega$ and Fourier expanded the periodic phase factor $e^{i\delta\phi_k(t)}=\sum_m\mathcal{F}_k^{(m)}e^{im\Omega t}$. The tunneling signal then consists of a series of Gaussian peaks centered at $V_g=\epsilon_{\mathrm{eff}}(k)+m\Omega$, which correspond to the renormalized energy band and the respective Floquet replicas. Furthermore, given that for finite $\sigma$ the duration of the probing pulse and the resolution of the energy band are inversely proportional, the evolution of the instantaneous dispersion $\epsilon(k,t)$ cannot be determined simultaneously with its spectral features to arbitrary precision, leading to the manifestation of the time-energy uncertainty principle.

Interestingly, the momentum dependence of the periodic phase factor imposes a selection rule on the Floquet replicas and their measured signal. Provided that  $\delta\phi_k^{(2m)}(t)\sim\cos(k)$ and $\delta\phi_k^{(2m+1)}(t)\sim \sin(k)$, the Fourier coefficients of $e^{i\delta\phi_k(t)}$ satisfy
\begin{align}
    \mathcal{F}_{k=\pm\pi/2}^{(2m)}=0, \hspace{3mm} \mathcal{F}_{k=0}^{(2m+1)}=0. 
\end{align}
Consequently, at $k=0$ odd (even) replicas experience a destructive (constructive) interference, while the opposite behavior occurs at $k=\pm\pi/2$. Therefore, this momentum dependent selection rule essentially controls the alternating pattern of bright and dark spectral features observed in Fig. \ref{fig:ED}.

\section{Rotating frame derivation}\label{sec:appendixB}

We present the derivation of (\ref{t-V chain rotating}), which is inspired by a similar calculation in the supplemental material of Ref.~\onlinecite{Anatoli2016Schrieffer-Wolff}. Starting from the Hamiltonian (\ref{NN}) and the time-dependent unitary operator (\ref{R for tV chain}), the Baker-Campbell-Hausdorff theorem \cite{Hall2015BCH, bonfiglioli2011BCHbook} gives 
\begin{align}
    \mathcal{R}(t)c_j \mathcal{R}^\dagger(t)&=e^{-itV\big(n_{j-1}+n_{j+1}-1\big)}c_j.
\end{align}
As a result, the interaction term is eliminated due to the identity  $H^{[\mathcal{R}]}=\mathcal{R}(t) H \mathcal{R}^\dagger(t)-i \mathcal{R}(t) \partial_t \mathcal{R}^\dagger(t)$, and the hopping term acquires an occupation-dependent phase
\begin{align}
    \mathcal{R}(t)c_j^\dagger c_{j+1}\mathcal{R}^\dagger(t)=c^\dagger_j\bigg[ e^{itV\big(n_{j-1}+n_{j+1} -n_{j}-n_{j+2}\big)}\bigg]c_{j+1}.
\end{align}
Using the relation $e^{i\alpha n_{j}}=(1-n_{j})+e^{i\alpha}n_{j}$ together with $c_j^\dagger n_j=n_{j+1}c_{j+1}=0$, we find
\begin{align}
        \mathcal{R}(t)c_j^\dagger c_{j+1}\mathcal{R}^\dagger(t)&=\bigg[(1-n_{j-1})+e^{+itV}n_{j-1}\bigg]
        \\ \nonumber
        &\hspace{.5cm}\bigg[(1-n_{j+2})+e^{-itV}n_{j+2}\bigg]c_j^\dagger c_{j+1}, 
\end{align}
leading, therefore, to the final form shown in (\ref{t-V chain rotating}), where the resonant and off-resonant domain wall channels become evident.

\section{Spectral function at finite temperature}\label{AppendixC}

In this Appendix we work in the atomic limit of (\ref{NN}), \textit{i.e.} $J=0$, which renders the Hamiltonian to become diagonal in occupation basis. As a consequence, the thermal averages in the grand-canonical ensemble reduce to classical correlations and, by exploiting transfer matrix techniques, the spectral function at finite temperature can be determined analytically. We begin by solving the Heisenberg equation 
\begin{align}
    i\partial_t c_j=[c_j,H]
    \Rightarrow c_j(t)=e^{-itV(n_{j-1}+n_{j+1}-1)}c_j,
\end{align}
from which the spectral function can be shown to be diagonal in real space
\begin{align} \label{Fourier of A}
    A_{ij}(\omega)
    &=\int \frac{dt}{2\pi}e^{i\omega t}\langle\{c_i(t), c_j\}\rangle_\beta
    \\
    &=\delta_{ij}\int \frac{dt}{2\pi}e^{i\omega t}\bigg\langle e^{-itV(n_{j-1}+n_{j+1}-1)}\bigg\rangle_\beta
\end{align}
and, therefore, momentum independent $A(k,\omega)=A(\omega)$. To evaluate the thermal correlation above, we define the transfer matrix 
\begin{align}
    \mathcal{T}&=e^{-\beta V\big(n_a-\frac{1}{2}\big)\big(n_b-\frac{1}{2}\big)}=e^{-\beta V/4}
    \left(
    \begin{array}{cc}
        1 & z \\
        z & 1
    \end{array}
    \right),
\end{align}
where $z=e^{\beta V/2}$, together with $\mathcal{D}(t)=\text{diag}\big(1, e^{-itV}\big)$. The Boltzmann weight then factorizes over bonds, such that $\mathcal{Z}=\text{Tr}(\mathcal{T}^L)$. Upon diagonalizing the transfer matrix $\mathcal{T}=\sum_{s=\pm}\lambda_s|s\rangle\langle s|$, where $\lambda\pm=e^{-\beta V/4}(1\pm z)$ and $|\pm\rangle=(1, \pm1)^T/\sqrt{2}$, and taking the thermodynamic limit, so that the $\lambda_+$ eigenvalue dominates over $\lambda_-$, we finally find
\begin{align} 
    C_\beta(t)&=\left\langle e^{-iVt\left(n_{j-1}+n_{j+1}-1\right)}\right\rangle_{\beta}
    \\ \nonumber
    &=
    \frac{e^{itV}}{\mathcal{Z}}\sum_{\{n\}}\left\langle 
    \!
    \{n\}
    \!
    \left|e^{-\beta H}e^{-itV(n_{j-1}+n_{j+1})}\right|
    \!
    \{n\}
    \!
    \right\rangle
    \\
    &=
    e^{itV}
    \lim_{L\rightarrow\infty}
    \frac{\text{Tr}\left(\mathcal{D}(t)\mathcal{T}^2\mathcal{D}(t)\mathcal{T}^{L-2}\right)}{\text{Tr}\left(\mathcal{T}^L\right)}
    \\
    &=
    \frac{e^{itV}e^{-\beta V/2}}{\lambda_+^2}\langle+|\mathcal{D}(t)\mathcal{T}^2\mathcal{D}(t)|+\rangle
    \\
    &=
    \frac{\left(1+z^{2}\right)\cos\left(Vt\right) + 2z}{\left(1+z\right)^{2}},
\end{align}
which according to (\ref{Fourier of A}) returns the spectral function (\ref{spectral function}) and provides, therefore, a reference point for the origin of the three energy bands observed in Fig. \ref{fig:temperature}.

\vspace{5cm}

\bibliography{ref}

\begin{thebibliography}{49}%
\makeatletter
\providecommand \@ifxundefined [1]{%
 \@ifx{#1\undefined}
}%
\providecommand \@ifnum [1]{%
 \ifnum #1\expandafter \@firstoftwo
 \else \expandafter \@secondoftwo
 \fi
}%
\providecommand \@ifx [1]{%
 \ifx #1\expandafter \@firstoftwo
 \else \expandafter \@secondoftwo
 \fi
}%
\providecommand \natexlab [1]{#1}%
\providecommand \enquote  [1]{``#1''}%
\providecommand \bibnamefont  [1]{#1}%
\providecommand \bibfnamefont [1]{#1}%
\providecommand \citenamefont [1]{#1}%
\providecommand \href@noop [0]{\@secondoftwo}%
\providecommand \href [0]{\begingroup \@sanitize@url \@href}%
\providecommand \@href[1]{\@@startlink{#1}\@@href}%
\providecommand \@@href[1]{\endgroup#1\@@endlink}%
\providecommand \@sanitize@url [0]{\catcode `\\12\catcode `\$12\catcode
  `\&12\catcode `\#12\catcode `\^12\catcode `\_12\catcode `\%12\relax}%
\providecommand \@@startlink[1]{}%
\providecommand \@@endlink[0]{}%
\providecommand \url  [0]{\begingroup\@sanitize@url \@url }%
\providecommand \@url [1]{\endgroup\@href {#1}{\urlprefix }}%
\providecommand \urlprefix  [0]{URL }%
\providecommand \Eprint [0]{\href }%
\providecommand \doibase [0]{https://doi.org/}%
\providecommand \selectlanguage [0]{\@gobble}%
\providecommand \bibinfo  [0]{\@secondoftwo}%
\providecommand \bibfield  [0]{\@secondoftwo}%
\providecommand \translation [1]{[#1]}%
\providecommand \BibitemOpen [0]{}%
\providecommand \bibitemStop [0]{}%
\providecommand \bibitemNoStop [0]{.\EOS\space}%
\providecommand \EOS [0]{\spacefactor3000\relax}%
\providecommand \BibitemShut  [1]{\csname bibitem#1\endcsname}%
\let\auto@bib@innerbib\@empty
\bibitem [{\citenamefont {Mitrano}\ \emph {et~al.}(2016)\citenamefont
  {Mitrano}, \citenamefont {Cantaluppi}, \citenamefont {Nicoletti},
  \citenamefont {Kaiser}, \citenamefont {Perucchi}, \citenamefont {Lupi},
  \citenamefont {Di~Pietro}, \citenamefont {Pontiroli}, \citenamefont {Ricco},
  \citenamefont {Clark} \emph {et~al.}}]{light-induced2016superconductivity}%
  \BibitemOpen
  \bibfield  {author} {\bibinfo {author} {\bibfnamefont {M.}~\bibnamefont
  {Mitrano}}, \bibinfo {author} {\bibfnamefont {A.}~\bibnamefont {Cantaluppi}},
  \bibinfo {author} {\bibfnamefont {D.}~\bibnamefont {Nicoletti}}, \bibinfo
  {author} {\bibfnamefont {S.}~\bibnamefont {Kaiser}}, \bibinfo {author}
  {\bibfnamefont {A.}~\bibnamefont {Perucchi}}, \bibinfo {author}
  {\bibfnamefont {S.}~\bibnamefont {Lupi}}, \bibinfo {author} {\bibfnamefont
  {P.}~\bibnamefont {Di~Pietro}}, \bibinfo {author} {\bibfnamefont
  {D.}~\bibnamefont {Pontiroli}}, \bibinfo {author} {\bibfnamefont
  {M.}~\bibnamefont {Ricco}}, \bibinfo {author} {\bibfnamefont {S.~R.}\
  \bibnamefont {Clark}}, \emph {et~al.},\ }\bibfield  {title} {\bibinfo {title}
  {Possible light-induced superconductivity in $\mathrm{K}_3\mathrm{C}_{60}$ at
  high temperature},\ }\href@noop {} {\bibfield  {journal} {\bibinfo  {journal}
  {Nature}\ }\textbf {\bibinfo {volume} {530}},\ \bibinfo {pages} {461}
  (\bibinfo {year} {2016})}\BibitemShut {NoStop}%
\bibitem [{\citenamefont {Claassen}\ \emph {et~al.}(2019)\citenamefont
  {Claassen}, \citenamefont {Kennes}, \citenamefont {Zingl}, \citenamefont
  {Sentef},\ and\ \citenamefont {Rubio}}]{claassen2019superconductor}%
  \BibitemOpen
  \bibfield  {author} {\bibinfo {author} {\bibfnamefont {M.}~\bibnamefont
  {Claassen}}, \bibinfo {author} {\bibfnamefont {D.~M.}\ \bibnamefont
  {Kennes}}, \bibinfo {author} {\bibfnamefont {M.}~\bibnamefont {Zingl}},
  \bibinfo {author} {\bibfnamefont {M.~A.}\ \bibnamefont {Sentef}},\ and\
  \bibinfo {author} {\bibfnamefont {A.}~\bibnamefont {Rubio}},\ }\bibfield
  {title} {\bibinfo {title} {Universal optical control of chiral
  superconductors and {M}ajorana modes},\ }\href@noop {} {\bibfield  {journal}
  {\bibinfo  {journal} {Nature Physics}\ }\textbf {\bibinfo {volume} {15}},\
  \bibinfo {pages} {766} (\bibinfo {year} {2019})}\BibitemShut {NoStop}%
\bibitem [{\citenamefont {Rudner}\ and\ \citenamefont
  {Lindner}(2020{\natexlab{a}})}]{rudner2020induced-topology}%
  \BibitemOpen
  \bibfield  {author} {\bibinfo {author} {\bibfnamefont {M.~S.}\ \bibnamefont
  {Rudner}}\ and\ \bibinfo {author} {\bibfnamefont {N.~H.}\ \bibnamefont
  {Lindner}},\ }\bibfield  {title} {\bibinfo {title} {Band structure
  engineering and non-equilibrium dynamics in {F}loquet topological
  insulators},\ }\href@noop {} {\bibfield  {journal} {\bibinfo  {journal}
  {Nature reviews physics}\ }\textbf {\bibinfo {volume} {2}},\ \bibinfo {pages}
  {229} (\bibinfo {year} {2020}{\natexlab{a}})}\BibitemShut {NoStop}%
\bibitem [{\citenamefont {Jotzu}\ \emph {et~al.}(2014)\citenamefont {Jotzu},
  \citenamefont {Messer}, \citenamefont {Desbuquois}, \citenamefont {Lebrat},
  \citenamefont {Uehlinger}, \citenamefont {Greif},\ and\ \citenamefont
  {Esslinger}}]{optical-lattice2014topological}%
  \BibitemOpen
  \bibfield  {author} {\bibinfo {author} {\bibfnamefont {G.}~\bibnamefont
  {Jotzu}}, \bibinfo {author} {\bibfnamefont {M.}~\bibnamefont {Messer}},
  \bibinfo {author} {\bibfnamefont {R.}~\bibnamefont {Desbuquois}}, \bibinfo
  {author} {\bibfnamefont {M.}~\bibnamefont {Lebrat}}, \bibinfo {author}
  {\bibfnamefont {T.}~\bibnamefont {Uehlinger}}, \bibinfo {author}
  {\bibfnamefont {D.}~\bibnamefont {Greif}},\ and\ \bibinfo {author}
  {\bibfnamefont {T.}~\bibnamefont {Esslinger}},\ }\bibfield  {title} {\bibinfo
  {title} {Experimental realization of the topological {H}aldane model with
  ultracold fermions},\ }\href@noop {} {\bibfield  {journal} {\bibinfo
  {journal} {Nature}\ }\textbf {\bibinfo {volume} {515}},\ \bibinfo {pages}
  {237} (\bibinfo {year} {2014})}\BibitemShut {NoStop}%
\bibitem [{\citenamefont {Dmytruk}\ and\ \citenamefont
  {Schir{\`o}}(2022)}]{topological2022controlling}%
  \BibitemOpen
  \bibfield  {author} {\bibinfo {author} {\bibfnamefont {O.}~\bibnamefont
  {Dmytruk}}\ and\ \bibinfo {author} {\bibfnamefont {M.}~\bibnamefont
  {Schir{\`o}}},\ }\bibfield  {title} {\bibinfo {title} {Controlling
  topological phases of matter with quantum light},\ }\href@noop {} {\bibfield
  {journal} {\bibinfo  {journal} {Communications Physics}\ }\textbf {\bibinfo
  {volume} {5}},\ \bibinfo {pages} {271} (\bibinfo {year} {2022})}\BibitemShut
  {NoStop}%
\bibitem [{\citenamefont {De~La~Torre}\ \emph {et~al.}(2021)\citenamefont
  {De~La~Torre}, \citenamefont {Kennes}, \citenamefont {Claassen},
  \citenamefont {Gerber}, \citenamefont {McIver},\ and\ \citenamefont
  {Sentef}}]{Alberto2021Colloquium}%
  \BibitemOpen
  \bibfield  {author} {\bibinfo {author} {\bibfnamefont {A.}~\bibnamefont
  {De~La~Torre}}, \bibinfo {author} {\bibfnamefont {D.~M.}\ \bibnamefont
  {Kennes}}, \bibinfo {author} {\bibfnamefont {M.}~\bibnamefont {Claassen}},
  \bibinfo {author} {\bibfnamefont {S.}~\bibnamefont {Gerber}}, \bibinfo
  {author} {\bibfnamefont {J.~W.}\ \bibnamefont {McIver}},\ and\ \bibinfo
  {author} {\bibfnamefont {M.~A.}\ \bibnamefont {Sentef}},\ }\bibfield  {title}
  {\bibinfo {title} {Colloquium: Nonthermal pathways to ultrafast control in
  quantum materials},\ }\href@noop {} {\bibfield  {journal} {\bibinfo
  {journal} {Reviews of Modern Physics}\ }\textbf {\bibinfo {volume} {93}},\
  \bibinfo {pages} {041002} (\bibinfo {year} {2021})}\BibitemShut {NoStop}%
\bibitem [{\citenamefont {Rudner}\ and\ \citenamefont
  {Lindner}(2020{\natexlab{b}})}]{2020FloquetHandbook}%
  \BibitemOpen
  \bibfield  {author} {\bibinfo {author} {\bibfnamefont {M.~S.}\ \bibnamefont
  {Rudner}}\ and\ \bibinfo {author} {\bibfnamefont {N.~H.}\ \bibnamefont
  {Lindner}},\ }\bibfield  {title} {\bibinfo {title} {The {F}loquet engineer's
  handbook},\ }\href@noop {} {\bibfield  {journal} {\bibinfo  {journal} {arXiv
  preprint arXiv:2003.08252}\ } (\bibinfo {year}
  {2020}{\natexlab{b}})}\BibitemShut {NoStop}%
\bibitem [{\citenamefont {Bukov}\ \emph {et~al.}(2015)\citenamefont {Bukov},
  \citenamefont {D'Alessio},\ and\ \citenamefont
  {Polkovnikov}}]{Anatoli2015High-frequency}%
  \BibitemOpen
  \bibfield  {author} {\bibinfo {author} {\bibfnamefont {M.}~\bibnamefont
  {Bukov}}, \bibinfo {author} {\bibfnamefont {L.}~\bibnamefont {D'Alessio}},\
  and\ \bibinfo {author} {\bibfnamefont {A.}~\bibnamefont {Polkovnikov}},\
  }\bibfield  {title} {\bibinfo {title} {Universal high-frequency behavior of
  periodically driven systems: from dynamical stabilization to {F}loquet
  engineering},\ }\href@noop {} {\bibfield  {journal} {\bibinfo  {journal}
  {Advances in Physics}\ }\textbf {\bibinfo {volume} {64}},\ \bibinfo {pages}
  {139} (\bibinfo {year} {2015})}\BibitemShut {NoStop}%
\bibitem [{\citenamefont {Oka}\ and\ \citenamefont
  {Kitamura}(2019)}]{oka2019FloquetEngineering}%
  \BibitemOpen
  \bibfield  {author} {\bibinfo {author} {\bibfnamefont {T.}~\bibnamefont
  {Oka}}\ and\ \bibinfo {author} {\bibfnamefont {S.}~\bibnamefont {Kitamura}},\
  }\bibfield  {title} {\bibinfo {title} {{F}loquet engineering of quantum
  materials},\ }\href@noop {} {\bibfield  {journal} {\bibinfo  {journal}
  {Annual Review of Condensed Matter Physics}\ }\textbf {\bibinfo {volume}
  {10}},\ \bibinfo {pages} {387} (\bibinfo {year} {2019})}\BibitemShut
  {NoStop}%
\bibitem [{\citenamefont {Weitenberg}\ and\ \citenamefont
  {Simonet}(2021)}]{weitenberg2021tailoring}%
  \BibitemOpen
  \bibfield  {author} {\bibinfo {author} {\bibfnamefont {C.}~\bibnamefont
  {Weitenberg}}\ and\ \bibinfo {author} {\bibfnamefont {J.}~\bibnamefont
  {Simonet}},\ }\bibfield  {title} {\bibinfo {title} {Tailoring quantum gases
  by {F}loquet engineering},\ }\href@noop {} {\bibfield  {journal} {\bibinfo
  {journal} {Nature Physics}\ }\textbf {\bibinfo {volume} {17}},\ \bibinfo
  {pages} {1342} (\bibinfo {year} {2021})}\BibitemShut {NoStop}%
\bibitem [{\citenamefont {Ito}\ \emph {et~al.}(2023)\citenamefont {Ito},
  \citenamefont {Sch{\"u}ler}, \citenamefont {Meierhofer}, \citenamefont
  {Schlauderer}, \citenamefont {Freudenstein}, \citenamefont {Reimann},
  \citenamefont {Afanasiev}, \citenamefont {Kokh}, \citenamefont
  {Tereshchenko}, \citenamefont {G{\"u}dde} \emph {et~al.}}]{exp1-2023}%
  \BibitemOpen
  \bibfield  {author} {\bibinfo {author} {\bibfnamefont {S.}~\bibnamefont
  {Ito}}, \bibinfo {author} {\bibfnamefont {M.}~\bibnamefont {Sch{\"u}ler}},
  \bibinfo {author} {\bibfnamefont {M.}~\bibnamefont {Meierhofer}}, \bibinfo
  {author} {\bibfnamefont {S.}~\bibnamefont {Schlauderer}}, \bibinfo {author}
  {\bibfnamefont {J.}~\bibnamefont {Freudenstein}}, \bibinfo {author}
  {\bibfnamefont {J.}~\bibnamefont {Reimann}}, \bibinfo {author} {\bibfnamefont
  {D.}~\bibnamefont {Afanasiev}}, \bibinfo {author} {\bibfnamefont
  {K.}~\bibnamefont {Kokh}}, \bibinfo {author} {\bibfnamefont {O.}~\bibnamefont
  {Tereshchenko}}, \bibinfo {author} {\bibfnamefont {J.}~\bibnamefont
  {G{\"u}dde}}, \emph {et~al.},\ }\bibfield  {title} {\bibinfo {title}
  {Build-up and dephasing of {F}loquet-{B}loch bands on subcycle timescales},\
  }\href@noop {} {\bibfield  {journal} {\bibinfo  {journal} {Nature}\ }\textbf
  {\bibinfo {volume} {616}},\ \bibinfo {pages} {696} (\bibinfo {year}
  {2023})}\BibitemShut {NoStop}%
\bibitem [{\citenamefont {Wang}\ \emph {et~al.}(2013)\citenamefont {Wang},
  \citenamefont {Steinberg}, \citenamefont {Jarillo-Herrero},\ and\
  \citenamefont {Gedik}}]{exp2-2013}%
  \BibitemOpen
  \bibfield  {author} {\bibinfo {author} {\bibfnamefont {Y.}~\bibnamefont
  {Wang}}, \bibinfo {author} {\bibfnamefont {H.}~\bibnamefont {Steinberg}},
  \bibinfo {author} {\bibfnamefont {P.}~\bibnamefont {Jarillo-Herrero}},\ and\
  \bibinfo {author} {\bibfnamefont {N.}~\bibnamefont {Gedik}},\ }\bibfield
  {title} {\bibinfo {title} {Observation of {F}loquet-{B}loch states on the
  surface of a topological insulator},\ }\href@noop {} {\bibfield  {journal}
  {\bibinfo  {journal} {Science}\ }\textbf {\bibinfo {volume} {342}},\ \bibinfo
  {pages} {453} (\bibinfo {year} {2013})}\BibitemShut {NoStop}%
\bibitem [{\citenamefont {Mahmood}\ \emph {et~al.}(2016)\citenamefont
  {Mahmood}, \citenamefont {Chan}, \citenamefont {Alpichshev}, \citenamefont
  {Gardner}, \citenamefont {Lee}, \citenamefont {Lee},\ and\ \citenamefont
  {Gedik}}]{exp3-2016}%
  \BibitemOpen
  \bibfield  {author} {\bibinfo {author} {\bibfnamefont {F.}~\bibnamefont
  {Mahmood}}, \bibinfo {author} {\bibfnamefont {C.-K.}\ \bibnamefont {Chan}},
  \bibinfo {author} {\bibfnamefont {Z.}~\bibnamefont {Alpichshev}}, \bibinfo
  {author} {\bibfnamefont {D.}~\bibnamefont {Gardner}}, \bibinfo {author}
  {\bibfnamefont {Y.}~\bibnamefont {Lee}}, \bibinfo {author} {\bibfnamefont
  {P.~A.}\ \bibnamefont {Lee}},\ and\ \bibinfo {author} {\bibfnamefont
  {N.}~\bibnamefont {Gedik}},\ }\bibfield  {title} {\bibinfo {title} {Selective
  scattering between {F}loquet-{B}loch and {V}olkov states in a topological
  insulator},\ }\href@noop {} {\bibfield  {journal} {\bibinfo  {journal}
  {Nature Physics}\ }\textbf {\bibinfo {volume} {12}},\ \bibinfo {pages} {306}
  (\bibinfo {year} {2016})}\BibitemShut {NoStop}%
\bibitem [{\citenamefont {Uhrig}\ \emph {et~al.}(2019)\citenamefont {Uhrig},
  \citenamefont {Kalthoff},\ and\ \citenamefont
  {Freericks}}]{uhrig2019positivity}%
  \BibitemOpen
  \bibfield  {author} {\bibinfo {author} {\bibfnamefont {G.~S.}\ \bibnamefont
  {Uhrig}}, \bibinfo {author} {\bibfnamefont {M.~H.}\ \bibnamefont
  {Kalthoff}},\ and\ \bibinfo {author} {\bibfnamefont {J.~K.}\ \bibnamefont
  {Freericks}},\ }\bibfield  {title} {\bibinfo {title} {Positivity of the
  spectral densities of retarded {F}loquet {G}reen functions},\ }\href@noop {}
  {\bibfield  {journal} {\bibinfo  {journal} {Physical Review Letters}\
  }\textbf {\bibinfo {volume} {122}},\ \bibinfo {pages} {130604} (\bibinfo
  {year} {2019})}\BibitemShut {NoStop}%
\bibitem [{\citenamefont {Zawadzki}\ and\ \citenamefont
  {Feiguin}(2019)}]{Krissia2019}%
  \BibitemOpen
  \bibfield  {author} {\bibinfo {author} {\bibfnamefont {K.}~\bibnamefont
  {Zawadzki}}\ and\ \bibinfo {author} {\bibfnamefont {A.~E.}\ \bibnamefont
  {Feiguin}},\ }\bibfield  {title} {\bibinfo {title} {Time- and
  momentum-resolved tunneling spectroscopy of pump-driven nonthermal
  excitations in {M}ott insulators},\ }\href@noop {} {\bibfield  {journal}
  {\bibinfo  {journal} {Physical Review B}\ }\textbf {\bibinfo {volume}
  {100}},\ \bibinfo {pages} {195124} (\bibinfo {year} {2019})}\BibitemShut
  {NoStop}%
\bibitem [{\citenamefont {Zawadzki}\ \emph {et~al.}(2020)\citenamefont
  {Zawadzki}, \citenamefont {Yang},\ and\ \citenamefont
  {Feiguin}}]{Krissia2020RIXS}%
  \BibitemOpen
  \bibfield  {author} {\bibinfo {author} {\bibfnamefont {K.}~\bibnamefont
  {Zawadzki}}, \bibinfo {author} {\bibfnamefont {L.}~\bibnamefont {Yang}},\
  and\ \bibinfo {author} {\bibfnamefont {A.~E.}\ \bibnamefont {Feiguin}},\
  }\bibfield  {title} {\bibinfo {title} {Time-dependent approach to inelastic
  scattering spectroscopies in and away from equilibrium: Beyond perturbation
  theory},\ }\href {https://doi.org/10.1103/PhysRevB.102.235141} {\bibfield
  {journal} {\bibinfo  {journal} {Phys. Rev. B}\ }\textbf {\bibinfo {volume}
  {102}},\ \bibinfo {pages} {235141} (\bibinfo {year} {2020})}\BibitemShut
  {NoStop}%
\bibitem [{\citenamefont {Zawadzki}\ \emph {et~al.}(2023)\citenamefont
  {Zawadzki}, \citenamefont {Nocera},\ and\ \citenamefont
  {Feiguin}}]{Krissia2023core-hole}%
  \BibitemOpen
  \bibfield  {author} {\bibinfo {author} {\bibfnamefont {K.}~\bibnamefont
  {Zawadzki}}, \bibinfo {author} {\bibfnamefont {A.}~\bibnamefont {Nocera}},\
  and\ \bibinfo {author} {\bibfnamefont {A.~E.}\ \bibnamefont {Feiguin}},\
  }\bibfield  {title} {\bibinfo {title} {A time-dependent momentum-resolved
  scattering approach to core-level spectroscopies},\ }\href
  {https://doi.org/10.21468/SciPostPhys.15.4.166} {\bibfield  {journal}
  {\bibinfo  {journal} {SciPost Phys.}\ }\textbf {\bibinfo {volume} {15}},\
  \bibinfo {pages} {166} (\bibinfo {year} {2023})}\BibitemShut {NoStop}%
\bibitem [{\citenamefont {Auslaender}\ \emph {et~al.}(2002)\citenamefont
  {Auslaender}, \citenamefont {Yacoby}, \citenamefont {de~Picciotto},
  \citenamefont {Baldwin}, \citenamefont {Pfeiffer},\ and\ \citenamefont
  {West}}]{auslaender2002}%
  \BibitemOpen
  \bibfield  {author} {\bibinfo {author} {\bibfnamefont {O.~M.}\ \bibnamefont
  {Auslaender}}, \bibinfo {author} {\bibfnamefont {A.}~\bibnamefont {Yacoby}},
  \bibinfo {author} {\bibfnamefont {R.}~\bibnamefont {de~Picciotto}}, \bibinfo
  {author} {\bibfnamefont {K.~W.}\ \bibnamefont {Baldwin}}, \bibinfo {author}
  {\bibfnamefont {L.~N.}\ \bibnamefont {Pfeiffer}},\ and\ \bibinfo {author}
  {\bibfnamefont {K.~W.}\ \bibnamefont {West}},\ }\bibfield  {title} {\bibinfo
  {title} {Tunneling spectroscopy of the elementary excitations in a
  one-dimensional wire},\ }\href {https://doi.org/10.1126/science.1066266}
  {\bibfield  {journal} {\bibinfo  {journal} {Science}\ }\textbf {\bibinfo
  {volume} {295}},\ \bibinfo {pages} {825} (\bibinfo {year}
  {2002})}\BibitemShut {NoStop}%
\bibitem [{\citenamefont {Auslaender}\ \emph {et~al.}(2005)\citenamefont
  {Auslaender}, \citenamefont {Steinberg}, \citenamefont {Yacoby},
  \citenamefont {Tserkovnyak}, \citenamefont {Halperin}, \citenamefont
  {Baldwin}, \citenamefont {Pfeiffer},\ and\ \citenamefont
  {West}}]{auslaender2005}%
  \BibitemOpen
  \bibfield  {author} {\bibinfo {author} {\bibfnamefont {O.~M.}\ \bibnamefont
  {Auslaender}}, \bibinfo {author} {\bibfnamefont {H.}~\bibnamefont
  {Steinberg}}, \bibinfo {author} {\bibfnamefont {A.}~\bibnamefont {Yacoby}},
  \bibinfo {author} {\bibfnamefont {Y.}~\bibnamefont {Tserkovnyak}}, \bibinfo
  {author} {\bibfnamefont {B.~I.}\ \bibnamefont {Halperin}}, \bibinfo {author}
  {\bibfnamefont {K.~W.}\ \bibnamefont {Baldwin}}, \bibinfo {author}
  {\bibfnamefont {L.~N.}\ \bibnamefont {Pfeiffer}},\ and\ \bibinfo {author}
  {\bibfnamefont {K.~W.}\ \bibnamefont {West}},\ }\bibfield  {title} {\bibinfo
  {title} {Spin-charge separation and localization in one dimension},\ }\href
  {https://doi.org/10.1126/science.1107821} {\bibfield  {journal} {\bibinfo
  {journal} {Science}\ }\textbf {\bibinfo {volume} {308}},\ \bibinfo {pages}
  {88} (\bibinfo {year} {2005})}\BibitemShut {NoStop}%
\bibitem [{\citenamefont {Inbar}\ \emph {et~al.}(2023)\citenamefont {Inbar},
  \citenamefont {Birkbeck}, \citenamefont {Xiao}, \citenamefont {Taniguchi},
  \citenamefont {Watanabe}, \citenamefont {Yan}, \citenamefont {Oreg},
  \citenamefont {Stern}, \citenamefont {Berg},\ and\ \citenamefont
  {Ilani}}]{Inbar2023}%
  \BibitemOpen
  \bibfield  {author} {\bibinfo {author} {\bibfnamefont {A.}~\bibnamefont
  {Inbar}}, \bibinfo {author} {\bibfnamefont {J.}~\bibnamefont {Birkbeck}},
  \bibinfo {author} {\bibfnamefont {J.}~\bibnamefont {Xiao}}, \bibinfo {author}
  {\bibfnamefont {T.}~\bibnamefont {Taniguchi}}, \bibinfo {author}
  {\bibfnamefont {K.}~\bibnamefont {Watanabe}}, \bibinfo {author}
  {\bibfnamefont {B.}~\bibnamefont {Yan}}, \bibinfo {author} {\bibfnamefont
  {Y.}~\bibnamefont {Oreg}}, \bibinfo {author} {\bibfnamefont {A.}~\bibnamefont
  {Stern}}, \bibinfo {author} {\bibfnamefont {E.}~\bibnamefont {Berg}},\ and\
  \bibinfo {author} {\bibfnamefont {S.}~\bibnamefont {Ilani}},\ }\bibfield
  {title} {\bibinfo {title} {The quantum twisting microscope},\ }\href
  {https://doi.org/10.1038/s41586-022-05685-y} {\bibfield  {journal} {\bibinfo
  {journal} {Nature}\ }\textbf {\bibinfo {volume} {614}},\ \bibinfo {pages}
  {682} (\bibinfo {year} {2023})}\BibitemShut {NoStop}%
\bibitem [{\citenamefont {Birkbeck}\ \emph {et~al.}(2025)\citenamefont
  {Birkbeck}, \citenamefont {Xiao}, \citenamefont {Inbar}, \citenamefont
  {Taniguchi}, \citenamefont {Watanabe}, \citenamefont {Berg}, \citenamefont
  {Glazman}, \citenamefont {Guinea}, \citenamefont {von Oppen},\ and\
  \citenamefont {Ilani}}]{Birkbeck2025}%
  \BibitemOpen
  \bibfield  {author} {\bibinfo {author} {\bibfnamefont {J.}~\bibnamefont
  {Birkbeck}}, \bibinfo {author} {\bibfnamefont {J.}~\bibnamefont {Xiao}},
  \bibinfo {author} {\bibfnamefont {A.}~\bibnamefont {Inbar}}, \bibinfo
  {author} {\bibfnamefont {T.}~\bibnamefont {Taniguchi}}, \bibinfo {author}
  {\bibfnamefont {K.}~\bibnamefont {Watanabe}}, \bibinfo {author}
  {\bibfnamefont {E.}~\bibnamefont {Berg}}, \bibinfo {author} {\bibfnamefont
  {L.}~\bibnamefont {Glazman}}, \bibinfo {author} {\bibfnamefont
  {F.}~\bibnamefont {Guinea}}, \bibinfo {author} {\bibfnamefont
  {F.}~\bibnamefont {von Oppen}},\ and\ \bibinfo {author} {\bibfnamefont
  {S.}~\bibnamefont {Ilani}},\ }\bibfield  {title} {\bibinfo {title} {Quantum
  twisting microscopy of phonons in twisted bilayer graphene},\ }\href
  {https://doi.org/10.1038/s41586-025-08881-8} {\bibfield  {journal} {\bibinfo
  {journal} {Nature}\ }\textbf {\bibinfo {volume} {641}},\ \bibinfo {pages}
  {345} (\bibinfo {year} {2025})}\BibitemShut {NoStop}%
\bibitem [{\citenamefont {Xiao}\ \emph {et~al.}(2026)\citenamefont {Xiao},
  \citenamefont {Inbar}, \citenamefont {Birkbeck}, \citenamefont {Gershon},
  \citenamefont {Zamir}, \citenamefont {Vituri}, \citenamefont {Taniguchi},
  \citenamefont {Watanabe}, \citenamefont {Berg},\ and\ \citenamefont
  {Ilani}}]{Xiao2026}%
  \BibitemOpen
  \bibfield  {author} {\bibinfo {author} {\bibfnamefont {J.}~\bibnamefont
  {Xiao}}, \bibinfo {author} {\bibfnamefont {A.}~\bibnamefont {Inbar}},
  \bibinfo {author} {\bibfnamefont {J.}~\bibnamefont {Birkbeck}}, \bibinfo
  {author} {\bibfnamefont {N.}~\bibnamefont {Gershon}}, \bibinfo {author}
  {\bibfnamefont {Y.}~\bibnamefont {Zamir}}, \bibinfo {author} {\bibfnamefont
  {Y.}~\bibnamefont {Vituri}}, \bibinfo {author} {\bibfnamefont
  {T.}~\bibnamefont {Taniguchi}}, \bibinfo {author} {\bibfnamefont
  {K.}~\bibnamefont {Watanabe}}, \bibinfo {author} {\bibfnamefont
  {E.}~\bibnamefont {Berg}},\ and\ \bibinfo {author} {\bibfnamefont
  {S.}~\bibnamefont {Ilani}},\ }\bibfield  {title} {\bibinfo {title} {Imaging
  the flat bands of magic-angle graphene reshaped by interactions},\ }\href
  {https://doi.org/10.1038/s41586-026-10378-x} {\bibfield  {journal} {\bibinfo
  {journal} {Nature}\ }\textbf {\bibinfo {volume} {653}},\ \bibinfo {pages}
  {68} (\bibinfo {year} {2026})}\BibitemShut {NoStop}%
\bibitem [{\citenamefont {Osterkorn}\ \emph {et~al.}(2023)\citenamefont
  {Osterkorn}, \citenamefont {Meyer},\ and\ \citenamefont
  {Manmana}}]{Germans2023InGap}%
  \BibitemOpen
  \bibfield  {author} {\bibinfo {author} {\bibfnamefont {A.}~\bibnamefont
  {Osterkorn}}, \bibinfo {author} {\bibfnamefont {C.}~\bibnamefont {Meyer}},\
  and\ \bibinfo {author} {\bibfnamefont {S.~R.}\ \bibnamefont {Manmana}},\
  }\bibfield  {title} {\bibinfo {title} {In-gap band formation in a
  periodically driven charge density wave insulator},\ }\href@noop {}
  {\bibfield  {journal} {\bibinfo  {journal} {Communications Physics}\ }\textbf
  {\bibinfo {volume} {6}},\ \bibinfo {pages} {245} (\bibinfo {year}
  {2023})}\BibitemShut {NoStop}%
\bibitem [{\citenamefont {Dunlap}\ and\ \citenamefont
  {Kenkre}(1986)}]{Dunlap1986dynamical}%
  \BibitemOpen
  \bibfield  {author} {\bibinfo {author} {\bibfnamefont {D.~H.}\ \bibnamefont
  {Dunlap}}\ and\ \bibinfo {author} {\bibfnamefont {V.~M.}\ \bibnamefont
  {Kenkre}},\ }\bibfield  {title} {\bibinfo {title} {Dynamic localization of a
  charged particle moving under the influence of an electric field},\ }\href
  {https://doi.org/10.1103/PhysRevB.34.3625} {\bibfield  {journal} {\bibinfo
  {journal} {Phys. Rev. B}\ }\textbf {\bibinfo {volume} {34}},\ \bibinfo
  {pages} {3625} (\bibinfo {year} {1986})}\BibitemShut {NoStop}%
\bibitem [{\citenamefont {White}(1992)}]{white1992DMRG}%
  \BibitemOpen
  \bibfield  {author} {\bibinfo {author} {\bibfnamefont {S.~R.}\ \bibnamefont
  {White}},\ }\bibfield  {title} {\bibinfo {title} {Density matrix formulation
  for quantum renormalization groups},\ }\href@noop {} {\bibfield  {journal}
  {\bibinfo  {journal} {Physical review letters}\ }\textbf {\bibinfo {volume}
  {69}},\ \bibinfo {pages} {2863} (\bibinfo {year} {1992})}\BibitemShut
  {NoStop}%
\bibitem [{\citenamefont {White}(1993)}]{white1993DMRG}%
  \BibitemOpen
  \bibfield  {author} {\bibinfo {author} {\bibfnamefont {S.~R.}\ \bibnamefont
  {White}},\ }\bibfield  {title} {\bibinfo {title} {Density-matrix algorithms
  for quantum renormalization groups},\ }\href
  {https://doi.org/10.1103/PhysRevB.48.10345} {\bibfield  {journal} {\bibinfo
  {journal} {Phys. Rev. B}\ }\textbf {\bibinfo {volume} {48}},\ \bibinfo
  {pages} {10345} (\bibinfo {year} {1993})}\BibitemShut {NoStop}%
\bibitem [{\citenamefont {Schollw{\"o}ck}(2005)}]{schollwock2005DMRG}%
  \BibitemOpen
  \bibfield  {author} {\bibinfo {author} {\bibfnamefont {U.}~\bibnamefont
  {Schollw{\"o}ck}},\ }\bibfield  {title} {\bibinfo {title} {The density-matrix
  renormalization group},\ }\href@noop {} {\bibfield  {journal} {\bibinfo
  {journal} {Reviews of modern physics}\ }\textbf {\bibinfo {volume} {77}},\
  \bibinfo {pages} {259} (\bibinfo {year} {2005})}\BibitemShut {NoStop}%
\bibitem [{\citenamefont {Schollw{\"o}ck}(2011)}]{schollwock2011DMRG}%
  \BibitemOpen
  \bibfield  {author} {\bibinfo {author} {\bibfnamefont {U.}~\bibnamefont
  {Schollw{\"o}ck}},\ }\bibfield  {title} {\bibinfo {title} {The density-matrix
  renormalization group in the age of matrix product states},\ }\href@noop {}
  {\bibfield  {journal} {\bibinfo  {journal} {Annals of physics}\ }\textbf
  {\bibinfo {volume} {326}},\ \bibinfo {pages} {96} (\bibinfo {year}
  {2011})}\BibitemShut {NoStop}%
\bibitem [{\citenamefont {Feiguin}(2011)}]{vietri}%
  \BibitemOpen
  \bibfield  {author} {\bibinfo {author} {\bibfnamefont {A.~E.}\ \bibnamefont
  {Feiguin}},\ }\bibfield  {title} {\bibinfo {title} {The density matrix
  renormalization group method and its time-dependent variants},\ }in\
  \href@noop {} {\emph {\bibinfo {booktitle} {XV Training Course in the Physics
  of Strongly Correlated Systems}}},\ Vol.\ \bibinfo {volume} {1419}\ (\bibinfo
   {publisher} {AIP Proceedings},\ \bibinfo {year} {2011})\ p.~\bibinfo {pages}
  {5}\BibitemShut {NoStop}%
\bibitem [{\citenamefont {Paeckel}\ \emph {et~al.}(2019)\citenamefont
  {Paeckel}, \citenamefont {Köhler}, \citenamefont {Swoboda}, \citenamefont
  {Manmana}, \citenamefont {Schollwöck},\ and\ \citenamefont
  {Hubig}}]{Paeckel2019}%
  \BibitemOpen
  \bibfield  {author} {\bibinfo {author} {\bibfnamefont {S.}~\bibnamefont
  {Paeckel}}, \bibinfo {author} {\bibfnamefont {T.}~\bibnamefont {Köhler}},
  \bibinfo {author} {\bibfnamefont {A.}~\bibnamefont {Swoboda}}, \bibinfo
  {author} {\bibfnamefont {S.~R.}\ \bibnamefont {Manmana}}, \bibinfo {author}
  {\bibfnamefont {U.}~\bibnamefont {Schollwöck}},\ and\ \bibinfo {author}
  {\bibfnamefont {C.}~\bibnamefont {Hubig}},\ }\bibfield  {title} {\bibinfo
  {title} {Time-evolution methods for matrix-product states},\ }\href
  {https://doi.org/https://doi.org/10.1016/j.aop.2019.167998} {\bibfield
  {journal} {\bibinfo  {journal} {Annals of Physics}\ }\textbf {\bibinfo
  {volume} {411}},\ \bibinfo {pages} {167998} (\bibinfo {year}
  {2019})}\BibitemShut {NoStop}%
\bibitem [{\citenamefont {Freericks}\ \emph {et~al.}(2009)\citenamefont
  {Freericks}, \citenamefont {Krishnamurthy},\ and\ \citenamefont
  {Pruschke}}]{freericks2009positivity}%
  \BibitemOpen
  \bibfield  {author} {\bibinfo {author} {\bibfnamefont {J.}~\bibnamefont
  {Freericks}}, \bibinfo {author} {\bibfnamefont {H.}~\bibnamefont
  {Krishnamurthy}},\ and\ \bibinfo {author} {\bibfnamefont {T.}~\bibnamefont
  {Pruschke}},\ }\bibfield  {title} {\bibinfo {title} {Theoretical description
  of time-resolved photoemission spectroscopy: application to pump-probe
  experiments},\ }\href@noop {} {\bibfield  {journal} {\bibinfo  {journal}
  {Physical review letters}\ }\textbf {\bibinfo {volume} {102}},\ \bibinfo
  {pages} {136401} (\bibinfo {year} {2009})}\BibitemShut {NoStop}%
\bibitem [{\citenamefont {Cohen}\ \emph {et~al.}(2014)\citenamefont {Cohen},
  \citenamefont {Gull}, \citenamefont {Reichman},\ and\ \citenamefont
  {Millis}}]{Cohen2014}%
  \BibitemOpen
  \bibfield  {author} {\bibinfo {author} {\bibfnamefont {G.}~\bibnamefont
  {Cohen}}, \bibinfo {author} {\bibfnamefont {E.}~\bibnamefont {Gull}},
  \bibinfo {author} {\bibfnamefont {D.~R.}\ \bibnamefont {Reichman}},\ and\
  \bibinfo {author} {\bibfnamefont {A.~J.}\ \bibnamefont {Millis}},\ }\bibfield
   {title} {\bibinfo {title} {Green's functions from real-time bold-line monte
  carlo calculations: Spectral properties of the nonequilibrium anderson
  impurity model},\ }\href {https://doi.org/10.1103/PhysRevLett.112.146802}
  {\bibfield  {journal} {\bibinfo  {journal} {Phys. Rev. Lett.}\ }\textbf
  {\bibinfo {volume} {112}},\ \bibinfo {pages} {146802} (\bibinfo {year}
  {2014})}\BibitemShut {NoStop}%
\bibitem [{\citenamefont {Hauschild}\ \emph {et~al.}(2024)\citenamefont
  {Hauschild}, \citenamefont {Unfried}, \citenamefont {Anand}, \citenamefont
  {Andrews}, \citenamefont {Bintz}, \citenamefont {Borla}, \citenamefont
  {Divic}, \citenamefont {Drescher}, \citenamefont {Geiger}, \citenamefont
  {Hefel}, \citenamefont {Hémery}, \citenamefont {Kadow}, \citenamefont
  {Kemp}, \citenamefont {Kirchner}, \citenamefont {Liu}, \citenamefont
  {Möller}, \citenamefont {Parker}, \citenamefont {Rader}, \citenamefont
  {Romen}, \citenamefont {Scalet}, \citenamefont {Schoonderwoerd},
  \citenamefont {Schulz}, \citenamefont {Soejima}, \citenamefont {Thoma},
  \citenamefont {Wu}, \citenamefont {Zechmann}, \citenamefont {Zweng},
  \citenamefont {Mong}, \citenamefont {Zaletel},\ and\ \citenamefont
  {Pollmann}}]{tenpy2024}%
  \BibitemOpen
  \bibfield  {author} {\bibinfo {author} {\bibfnamefont {J.}~\bibnamefont
  {Hauschild}}, \bibinfo {author} {\bibfnamefont {J.}~\bibnamefont {Unfried}},
  \bibinfo {author} {\bibfnamefont {S.}~\bibnamefont {Anand}}, \bibinfo
  {author} {\bibfnamefont {B.}~\bibnamefont {Andrews}}, \bibinfo {author}
  {\bibfnamefont {M.}~\bibnamefont {Bintz}}, \bibinfo {author} {\bibfnamefont
  {U.}~\bibnamefont {Borla}}, \bibinfo {author} {\bibfnamefont
  {S.}~\bibnamefont {Divic}}, \bibinfo {author} {\bibfnamefont
  {M.}~\bibnamefont {Drescher}}, \bibinfo {author} {\bibfnamefont
  {J.}~\bibnamefont {Geiger}}, \bibinfo {author} {\bibfnamefont
  {M.}~\bibnamefont {Hefel}}, \bibinfo {author} {\bibfnamefont
  {K.}~\bibnamefont {Hémery}}, \bibinfo {author} {\bibfnamefont
  {W.}~\bibnamefont {Kadow}}, \bibinfo {author} {\bibfnamefont
  {J.}~\bibnamefont {Kemp}}, \bibinfo {author} {\bibfnamefont {N.}~\bibnamefont
  {Kirchner}}, \bibinfo {author} {\bibfnamefont {V.~S.}\ \bibnamefont {Liu}},
  \bibinfo {author} {\bibfnamefont {G.}~\bibnamefont {Möller}}, \bibinfo
  {author} {\bibfnamefont {D.}~\bibnamefont {Parker}}, \bibinfo {author}
  {\bibfnamefont {M.}~\bibnamefont {Rader}}, \bibinfo {author} {\bibfnamefont
  {A.}~\bibnamefont {Romen}}, \bibinfo {author} {\bibfnamefont
  {S.}~\bibnamefont {Scalet}}, \bibinfo {author} {\bibfnamefont
  {L.}~\bibnamefont {Schoonderwoerd}}, \bibinfo {author} {\bibfnamefont
  {M.}~\bibnamefont {Schulz}}, \bibinfo {author} {\bibfnamefont
  {T.}~\bibnamefont {Soejima}}, \bibinfo {author} {\bibfnamefont
  {P.}~\bibnamefont {Thoma}}, \bibinfo {author} {\bibfnamefont
  {Y.}~\bibnamefont {Wu}}, \bibinfo {author} {\bibfnamefont {P.}~\bibnamefont
  {Zechmann}}, \bibinfo {author} {\bibfnamefont {L.}~\bibnamefont {Zweng}},
  \bibinfo {author} {\bibfnamefont {R.~S.~K.}\ \bibnamefont {Mong}}, \bibinfo
  {author} {\bibfnamefont {M.~P.}\ \bibnamefont {Zaletel}},\ and\ \bibinfo
  {author} {\bibfnamefont {F.}~\bibnamefont {Pollmann}},\ }\bibfield  {title}
  {\bibinfo {title} {{Tensor network Python (TeNPy) version 1}},\ }\href
  {https://doi.org/10.21468/SciPostPhysCodeb.41} {\bibfield  {journal}
  {\bibinfo  {journal} {SciPost Phys. Codebases}\ ,\ \bibinfo {pages} {41}}
  (\bibinfo {year} {2024})}\BibitemShut {NoStop}%
\bibitem [{\citenamefont {Tsuji}\ \emph {et~al.}(2011)\citenamefont {Tsuji},
  \citenamefont {Oka}, \citenamefont {Werner},\ and\ \citenamefont
  {Aoki}}]{tsuji2011dynamical}%
  \BibitemOpen
  \bibfield  {author} {\bibinfo {author} {\bibfnamefont {N.}~\bibnamefont
  {Tsuji}}, \bibinfo {author} {\bibfnamefont {T.}~\bibnamefont {Oka}}, \bibinfo
  {author} {\bibfnamefont {P.}~\bibnamefont {Werner}},\ and\ \bibinfo {author}
  {\bibfnamefont {H.}~\bibnamefont {Aoki}},\ }\bibfield  {title} {\bibinfo
  {title} {Dynamical band flipping in fermionic lattice systems: An
  {AC}-field-driven change of the interaction from repulsive to attractive},\
  }\href@noop {} {\bibfield  {journal} {\bibinfo  {journal} {Physical Review
  Letters}\ }\textbf {\bibinfo {volume} {106}},\ \bibinfo {pages} {236401}
  (\bibinfo {year} {2011})}\BibitemShut {NoStop}%
\bibitem [{\citenamefont {Ikeda}\ and\ \citenamefont
  {Polkovnikov}(2021)}]{Anatoli2021fermi-golden-rule}%
  \BibitemOpen
  \bibfield  {author} {\bibinfo {author} {\bibfnamefont {T.~N.}\ \bibnamefont
  {Ikeda}}\ and\ \bibinfo {author} {\bibfnamefont {A.}~\bibnamefont
  {Polkovnikov}},\ }\bibfield  {title} {\bibinfo {title} {Fermi's golden rule
  for heating in strongly driven {F}loquet systems},\ }\href@noop {} {\bibfield
   {journal} {\bibinfo  {journal} {Physical Review B}\ }\textbf {\bibinfo
  {volume} {104}},\ \bibinfo {pages} {134308} (\bibinfo {year}
  {2021})}\BibitemShut {NoStop}%
\bibitem [{\citenamefont {Cupo}\ \emph {et~al.}(2025)\citenamefont {Cupo},
  \citenamefont {Cheng}, \citenamefont {Ramanathan},\ and\ \citenamefont
  {Viola}}]{cupo2025Dartmounth}%
  \BibitemOpen
  \bibfield  {author} {\bibinfo {author} {\bibfnamefont {A.}~\bibnamefont
  {Cupo}}, \bibinfo {author} {\bibfnamefont {H.-P.}\ \bibnamefont {Cheng}},
  \bibinfo {author} {\bibfnamefont {C.}~\bibnamefont {Ramanathan}},\ and\
  \bibinfo {author} {\bibfnamefont {L.}~\bibnamefont {Viola}},\ }\bibfield
  {title} {\bibinfo {title} {Generation of ultrahigh anomalous {H}all
  conductivities via optimally prepared topological {F}loquet states},\
  }\href@noop {} {\bibfield  {journal} {\bibinfo  {journal} {arXiv preprint
  arXiv:2511.19843}\ } (\bibinfo {year} {2025})}\BibitemShut {NoStop}%
\bibitem [{\citenamefont {Bukov}\ \emph {et~al.}(2016)\citenamefont {Bukov},
  \citenamefont {Kolodrubetz},\ and\ \citenamefont
  {Polkovnikov}}]{Anatoli2016Schrieffer-Wolff}%
  \BibitemOpen
  \bibfield  {author} {\bibinfo {author} {\bibfnamefont {M.}~\bibnamefont
  {Bukov}}, \bibinfo {author} {\bibfnamefont {M.}~\bibnamefont {Kolodrubetz}},\
  and\ \bibinfo {author} {\bibfnamefont {A.}~\bibnamefont {Polkovnikov}},\
  }\bibfield  {title} {\bibinfo {title} {{S}chrieffer-{W}olff transformation
  for periodically driven systems: Strongly correlated systems with artificial
  gauge fields},\ }\href@noop {} {\bibfield  {journal} {\bibinfo  {journal}
  {Physical review letters}\ }\textbf {\bibinfo {volume} {116}},\ \bibinfo
  {pages} {125301} (\bibinfo {year} {2016})}\BibitemShut {NoStop}%
\bibitem [{\citenamefont {Kumar}\ and\ \citenamefont
  {Lin}(2021)}]{kumar2021inducing}%
  \BibitemOpen
  \bibfield  {author} {\bibinfo {author} {\bibfnamefont {U.}~\bibnamefont
  {Kumar}}\ and\ \bibinfo {author} {\bibfnamefont {S.-Z.}\ \bibnamefont
  {Lin}},\ }\bibfield  {title} {\bibinfo {title} {Inducing and controlling
  superconductivity in the {H}ubbard honeycomb model using an electromagnetic
  drive},\ }\href@noop {} {\bibfield  {journal} {\bibinfo  {journal} {Physical
  Review B}\ }\textbf {\bibinfo {volume} {103}},\ \bibinfo {pages} {064508}
  (\bibinfo {year} {2021})}\BibitemShut {NoStop}%
\bibitem [{\citenamefont {Essler}\ \emph {et~al.}(2005)\citenamefont {Essler},
  \citenamefont {Frahm}, \citenamefont {G{\"o}hmann}, \citenamefont
  {Kl{\"u}mper},\ and\ \citenamefont {Korepin}}]{2005Hubbard-book}%
  \BibitemOpen
  \bibfield  {author} {\bibinfo {author} {\bibfnamefont {F.~H.}\ \bibnamefont
  {Essler}}, \bibinfo {author} {\bibfnamefont {H.}~\bibnamefont {Frahm}},
  \bibinfo {author} {\bibfnamefont {F.}~\bibnamefont {G{\"o}hmann}}, \bibinfo
  {author} {\bibfnamefont {A.}~\bibnamefont {Kl{\"u}mper}},\ and\ \bibinfo
  {author} {\bibfnamefont {V.~E.}\ \bibnamefont {Korepin}},\ }\href@noop {}
  {\emph {\bibinfo {title} {The one-dimensional Hubbard model}}},\
  Vol.~\bibinfo {volume} {10}\ (\bibinfo  {publisher} {Cambridge University
  Press Cambridge},\ \bibinfo {year} {2005})\BibitemShut {NoStop}%
\bibitem [{\citenamefont {Haldane}(1981)}]{Haldane1981luttinger}%
  \BibitemOpen
  \bibfield  {author} {\bibinfo {author} {\bibfnamefont {F.~D.~M.}\
  \bibnamefont {Haldane}},\ }\bibfield  {title} {\bibinfo {title}
  {``{L}uttinger liquid theory'' of one-dimensional quantum fluids. {I}.
  {P}roperties of the {L}uttinger model and their extension to the general 1{D}
  interacting spinless {F}ermi gas},\ }\href@noop {} {\bibfield  {journal}
  {\bibinfo  {journal} {Journal of Physics C: Solid State Physics}\ }\textbf
  {\bibinfo {volume} {14}},\ \bibinfo {pages} {2585} (\bibinfo {year}
  {1981})}\BibitemShut {NoStop}%
\bibitem [{\citenamefont {Giamarchi}(2004)}]{giamarchi2004quantum}%
  \BibitemOpen
  \bibfield  {author} {\bibinfo {author} {\bibfnamefont {T.}~\bibnamefont
  {Giamarchi}},\ }\href@noop {} {\emph {\bibinfo {title} {Quantum physics in
  one dimension}}},\ Vol.\ \bibinfo {volume} {121}\ (\bibinfo  {publisher}
  {Clarendon Oxford},\ \bibinfo {year} {2004})\BibitemShut {NoStop}%
\bibitem [{\citenamefont {Nocera}\ \emph {et~al.}(2018)\citenamefont {Nocera},
  \citenamefont {Essler},\ and\ \citenamefont
  {Feiguin}}]{nocera2018finite-temperature}%
  \BibitemOpen
  \bibfield  {author} {\bibinfo {author} {\bibfnamefont {A.}~\bibnamefont
  {Nocera}}, \bibinfo {author} {\bibfnamefont {F.~H.}\ \bibnamefont {Essler}},\
  and\ \bibinfo {author} {\bibfnamefont {A.~E.}\ \bibnamefont {Feiguin}},\
  }\bibfield  {title} {\bibinfo {title} {Finite-temperature dynamics of the
  {M}ott insulating {H}ubbard chain},\ }\href@noop {} {\bibfield  {journal}
  {\bibinfo  {journal} {Physical Review B}\ }\textbf {\bibinfo {volume} {97}},\
  \bibinfo {pages} {045146} (\bibinfo {year} {2018})}\BibitemShut {NoStop}%
\bibitem [{\citenamefont {Eskes}\ \emph {et~al.}(1991)\citenamefont {Eskes},
  \citenamefont {Meinders},\ and\ \citenamefont {Sawatzky}}]{Eskes1991}%
  \BibitemOpen
  \bibfield  {author} {\bibinfo {author} {\bibfnamefont {H.}~\bibnamefont
  {Eskes}}, \bibinfo {author} {\bibfnamefont {M.~B.~J.}\ \bibnamefont
  {Meinders}},\ and\ \bibinfo {author} {\bibfnamefont {G.~A.}\ \bibnamefont
  {Sawatzky}},\ }\bibfield  {title} {\bibinfo {title} {Anomalous transfer of
  spectral weight in doped strongly correlated systems},\ }\href
  {https://doi.org/10.1103/PhysRevLett.67.1035} {\bibfield  {journal} {\bibinfo
   {journal} {Phys. Rev. Lett.}\ }\textbf {\bibinfo {volume} {67}},\ \bibinfo
  {pages} {1035} (\bibinfo {year} {1991})}\BibitemShut {NoStop}%
\bibitem [{\citenamefont {Feiguin}\ and\ \citenamefont
  {White}(2005)}]{feiguin2005a}%
  \BibitemOpen
  \bibfield  {author} {\bibinfo {author} {\bibfnamefont {A.}~\bibnamefont
  {Feiguin}}\ and\ \bibinfo {author} {\bibfnamefont {S.}~\bibnamefont
  {White}},\ }\bibfield  {title} {\bibinfo {title} {Finite-temperature density
  matrix renormalization using an enlarged {H}ilbert space},\ }\href
  {https://doi.org/https://doi.org/10.1103/PhysRevB.72.220401} {\bibfield
  {journal} {\bibinfo  {journal} {Phys. Rev. B}\ }\textbf {\bibinfo {volume}
  {72}},\ \bibinfo {pages} {220401} (\bibinfo {year} {2005})}\BibitemShut
  {NoStop}%
\bibitem [{\citenamefont {Karrasch}\ \emph {et~al.}(2013)\citenamefont
  {Karrasch}, \citenamefont {Bardarson},\ and\ \citenamefont
  {Moore}}]{Karrasch2013}%
  \BibitemOpen
  \bibfield  {author} {\bibinfo {author} {\bibfnamefont {C.}~\bibnamefont
  {Karrasch}}, \bibinfo {author} {\bibfnamefont {J.~H.}\ \bibnamefont
  {Bardarson}},\ and\ \bibinfo {author} {\bibfnamefont {J.~E.}\ \bibnamefont
  {Moore}},\ }\bibfield  {title} {\bibinfo {title} {Reducing the numerical
  effort of finite-temperature density matrix renormalization group
  calculations},\ }\href {http://stacks.iop.org/1367-2630/15/i=8/a=083031}
  {\bibfield  {journal} {\bibinfo  {journal} {New Journal of Physics}\ }\textbf
  {\bibinfo {volume} {15}},\ \bibinfo {pages} {083031} (\bibinfo {year}
  {2013})}\BibitemShut {NoStop}%
\bibitem [{\citenamefont {Queiroz~Silveira}\ and\ \citenamefont
  {Feiguin}(2026)}]{Queiroz-2026}%
  \BibitemOpen
  \bibfield  {author} {\bibinfo {author} {\bibfnamefont {L.}~\bibnamefont
  {Queiroz~Silveira}}\ and\ \bibinfo {author} {\bibfnamefont {A.}~\bibnamefont
  {Feiguin}},\ }\bibfield  {title} {\bibinfo {title} {Time and {M}omentum
  {R}esolved {T}unneling {S}pectroscopy of {F}loquet dynamics},\ }\href
  {https://doi.org/10.5281/zenodo.24697623} {10.5281/zenodo.24697623} (\bibinfo
  {year} {2026})\BibitemShut {NoStop}%
\bibitem [{\citenamefont {Queiroz~Silveira}(2026)}]{Queiroz-2026-github}%
  \BibitemOpen
  \bibfield  {author} {\bibinfo {author} {\bibfnamefont {L.}~\bibnamefont
  {Queiroz~Silveira}},\ }\href@noop {} {\bibinfo {title} {Time-resolved floquet
  spectroscopy}},\ \bibinfo {howpublished}
  {\url{https://github.com/Lucas-Q-Silveira/Time-Resolved-Floquet-Spectroscopy}}
  (\bibinfo {year} {2026}),\ \bibinfo {note} {gitHub repository}\BibitemShut
  {NoStop}%
\bibitem [{\citenamefont {Hall}(2015)}]{Hall2015BCH}%
  \BibitemOpen
  \bibfield  {author} {\bibinfo {author} {\bibfnamefont {B.}~\bibnamefont
  {Hall}},\ }\bibinfo {title} {The {B}aker-{C}ampbell-{H}ausdorff formula and
  its consequences},\ in\ \href {https://doi.org/10.1007/978-3-319-13467-3_5}
  {\emph {\bibinfo {booktitle} {Lie Groups, Lie Algebras, and Representations:
  An Elementary Introduction}}}\ (\bibinfo  {publisher} {Springer International
  Publishing},\ \bibinfo {address} {Cham},\ \bibinfo {year} {2015})\ pp.\
  \bibinfo {pages} {109--137}\BibitemShut {NoStop}%
\bibitem [{\citenamefont {Bonfiglioli}\ and\ \citenamefont
  {Fulci}(2011)}]{bonfiglioli2011BCHbook}%
  \BibitemOpen
  \bibfield  {author} {\bibinfo {author} {\bibfnamefont {A.}~\bibnamefont
  {Bonfiglioli}}\ and\ \bibinfo {author} {\bibfnamefont {R.}~\bibnamefont
  {Fulci}},\ }\href@noop {} {\emph {\bibinfo {title} {Topics in noncommutative
  algebra: the theorem of {C}ampbell, {B}aker, {H}ausdorff and {D}ynkin}}}\
  (\bibinfo  {publisher} {Springer},\ \bibinfo {year} {2011})\BibitemShut
  {NoStop}%
\end{thebibliography}%

\end{document}